\documentclass[twocolumn]{aastex631}
\usepackage{color} 
\usepackage{tabularx} 
\usepackage{epsfig}
\usepackage{amsmath} 
\usepackage{amssymb} 
\usepackage{graphicx}
\usepackage{enumitem}
\usepackage{multirow}
\usepackage{cprotect}
\usepackage{natbib}
\setlist{nosep}
\usepackage{url}
\definecolor{purp}{HTML}{8904B1}

\let\oldhat\hat
\renewcommand{\hat}[1]{\oldhat{\mathbf{#1}}}

\let\oldwidehat\widehat
\renewcommand{\widehat}[1]{\oldwidehat{\mathbf{#1}}}

\newcommand\fe[3]{$#1^{+#2}_{-#3}$}

\newcommand{\griz}{\textsc{Grizli}}

\newcommand{\targ}{COBRA1411+3415}

\shortauthors{Watson et al.}
\submitjournal{ApJ}
\accepted{March 21, 2025}

\begin{document}
\defcitealias{GM2019}{GM19}
\defcitealias{PM2017}{PM17}

\title{HST Grism Observations of a $z\sim1.8$ Cluster Candidate \\ from the Clusters Occupied by Bent Radio AGN (COBRA) Survey}

\correspondingauthor{C. B. Watson}
\author[0000-0001-8456-4142]{Courtney B. Watson}
\affiliation{Institute for Astrophysical Research and Department of Astronomy, Boston University, Boston, MA 02215, USA}
\affiliation{Center for Astrophysics $|$ Harvard \& Smithsonian, 60 Garden Street, Cambridge, MA 02138, USA}
\email{cbwatson@bu.edu}

\author[0000-0002-0485-6573]{Elizabeth L. Blanton}
\affiliation{Institute for Astrophysical Research and Department of Astronomy, Boston University, Boston, MA 02215, USA}
% \email{eblanton@bu.edu}

\author[0000-0001-5160-6713]{Emmet Golden-Marx}
\affiliation{INAF - Osservatorio astronomico di Padova, Vicolo Osservatorio 5, 35122 Padova, Italy}
% \email{emmetgm@bu.edu}

\author[0000-0002-3993-0745]{Matthew L. N. Ashby}
\affiliation{Center for Astrophysics $|$ Harvard \& Smithsonian, 60 Garden Street, Cambridge, MA 02138, USA}
% \email{mashby@cfa.harvard.edu}

\author[0000-0002-3984-4337]{Scott W. Randall}
\affiliation{Center for Astrophysics $|$ Harvard \& Smithsonian, 60 Garden Street, Cambridge, MA 02138, USA}
% \email{srandall@cfa.harvard.edu}

\author[0009-0003-5349-6994]{J. D. Wing}
\affiliation{Center for Astrophysics $|$ Harvard \& Smithsonian, 60 Garden Street, Cambridge, MA 02138, USA}
% \email{jwing@cfa.harvard.edu}

\author[0000-0003-3646-3472]{E. M. Douglass}
\affiliation{Farmingdale State College—SUNY, 2350 Broadhollow Rd., Farmingdale, NY 11735, USA}
% \email{douglae@farmingdale.edu}

\begin{abstract}

We present new Hubble Space Telescope/Wide Field Camera 3 G141 grism observations for \targ, originally identified as a high-redshift cluster candidate in the Clusters Occupied by Bent Radio AGN (COBRA) survey using radio, infrared, and optical data. We spectroscopically identify seven cluster members within a 0.5 Mpc radius with grism redshifts in the range $1.8006 \leq z_{grism} \leq 1.8175$, consistent with \targ\ being a high-redshift cluster with a mean redshift of $\langle z_{grism}\rangle = 1.8106 \pm 0.0006$. The detection of seven galaxies within this small redshift range is significant above the background distribution of galaxies at the level of 5$\sigma$. The line-of-sight velocity dispersion of the cluster is found to be $\sigma_{\parallel} =$ \fe{701}{347}{138} km/s with a virial mass of $M_{200} \approx$ \fe{2.2}{3.3}{1.3}$\times 10^{14}$ M$_{\odot}$. However, the mass may be lower if the cluster is still in formation. In projected phase-space, we also identify two possible infalling members of \targ\ and two additional structures at $z\sim 1.73$ and $z\sim 1.88$. The similar spatial distributions and small projected separation from the main cluster suggest they could be a part of the same large-scale filament and together may form a protocluster system that could eventually merge to form a single, massive cluster. \targ\ is the highest redshift cluster to be spectroscopically confirmed using a bent, double-lobed radio source as a cluster tracer.
\end{abstract}

\section{Introduction}

The study of galaxy clusters is a cornerstone in our understanding of the formation and evolution of large-scale structures. As the most massive ($\sim 10^{13}-10^{15}$ M$_{\odot}$) virialized systems in the universe, galaxy clusters, comprised of tens to thousands of members, are ideal laboratories for investigating galaxy formation and evolution, and the effects of the cluster environment on these processes. Extensive identification and analysis of clusters at redshifts $<0.3$ has greatly enriched our knowledge of the low-$z$ universe. However, beyond $z>1.5$ the number of confirmed clusters drops significantly \citep{Papovich2010, Chiaberge2010, Noirot2016, Watson2019, Wen2021}. The identification, confirmation, and analysis of high-z clusters is vital for furthering our understanding of large-scale structure formation and evolution.

Galaxy clusters are detected using various methods, each with its own advantages and limitations. One common method is through X-ray observations of intracluster medium (ICM) emission, although this is biased towards more massive and lower redshift clusters, as the ICM of high-z clusters may be too faint to be detected in typical integration times implemented by X-ray observations, and cool core systems. Observations from eROSITA are greatly increasing the number of X-ray clusters found in surveys, extending above z=1 \citep{Bulbul2024}. 

Clusters can also be identified by their constituent galaxy members. This can be done using a variety of photometric or spectroscopic methods. The earliest efforts of cluster detection relied solely on optical measurements of galaxy overdensities \citep{Zwicky1950, Abell1958, Zwicky1961, Abell1962, Abell1989}, however, these surveys suffer from projection effects and are biased to low-redshift galaxies. 

Later studies utilized the fact that at $z>0.5$, the optical emission from a galaxy is shifted to infrared (IR) wavelengths, allowing for additional cluster detections based on IR galaxy overdensities \citep{Gladders2000, Eisenhardt2008, Zeimann2012, Stanford2012, Rettura2014, PM2017}. Another IR method utilizes a galaxy's 3.6 \micron\ $-$ 4.5 \micron\ colors. \cite{Papovich2008} showed that 90\% of galaxies at redshifts greater than 1.3 can be identified using a 3.6 \micron\ $-$ 4.5 \micron\ color cut. This method has led to the identification of many high-z clusters \citep{Papovich2010, Muzzin2013, Wylezalek2013, GM2019}. 

Another method of cluster detection involves measuring distortions in the cosmic microwave background (CMB) caused by the Sunyaev-Zel’Dovich (SZ) effect \citep{Sunyaev1972}. The SZ effect can be used to identify clusters at all redshifts, as the strength of the observed distortion is not dependent on the cluster redshift. However, this technique will preferentially detect more massive clusters as the observed SZ effect is proportional to the mass of the cluster causing the distortion \citep{Staniszewski2009,Bleem2015,Bleem2020}.

Recently, techniques using active galactic nuclei (AGN) with associated radio emission have shown promise for identifying cluster candidates. Surveys such as Clusters Around Radio-Loud AGN (CARLA) \citep{Galametz2012, Wylezalek2014, Cooke2015, Cooke2016, Noirot2018} and Clusters Occupied by Bent Radio AGN (COBRA) \citep{Wing2011, PM2017, GM2019, Golden-Marx2021, Golden-Marx2023} have identified many high-z clusters by targeting radio-loud AGN. CARLA targets AGN based on radio power, searching around powerful radio sources regardless of the radio morphology. Another study by \cite{Castignani2014} was successful in identifying regions of overdensities surrounding low-power radio sources.

COBRA specifically targets bent, double-lobed radio sources associated with AGN, regardless of radio power, as they directly probe the interaction of the radio source with the surrounding intracluster medium (ICM). The radio lobes of these sources are likely bent by the relative motion between the radio host galaxy and surrounding extended gas, such as the intracluster medium (ICM) or intragroup medium (IGM), making bent radio sources signposts for clusters and groups. \cite{Wing2011} show that $\approx$60-80\% of bent radio AGN, vs $\approx$20-40\% straight-lobed sources, are found in rich clusters or groups up to $z\approx 0.5$. Bent, double-lobed radio sources can be found in clusters with a wide range of redshifts, masses, and dynamical states \citep{Blanton2000, Blanton2001a, Blanton2003a, Miley2008, Douglass2008, Douglass2011,Douglass2018, Wing2011, Wing2013, PM2013, PM2017, ODea2023, Watson2023}, making them ideal for identifying a wide variety of high-z cluster candidates for follow-up observations.

\subsection{COBRA Target Selection}

The COBRA sample of bent radio AGN was initially curated by \cite{Wing2011} using the Very Large Array (VLA) Faint Images of the Radio Sky at Twenty-centimeters (FIRST) survey \citep{Becker1995} cross-correlated with the Sloan Digital Sky Survey Data Release 7 (SDSS-DR7; \cite{York2000,Abazajian_2009}) for detection of optical counterparts. The sources with known optical counterparts, which were found to lie at mostly $z<0.7$ based on the limits of SDSS, constitute the low-z portion of COBRA, where \cite{Wing2011} identified clusters based on the optical richness of the fields surrounding the radio sources. Quiescent host galaxies were generally not detected in the SDSS at redshifts $>0.7$; however, radio sources that have luminous quasars as hosts were detected in SDSS beyond $z=0.7$, typically with both photometry and spectroscopy available. 

Building on this work, \cite{PM2017} analyzed follow-up 3.6 \micron\ and 4.5 \micron\ Spitzer observations (Cycle 11; PI E Blanton) of the fields surrounding COBRA sources that either did not have an SDSS optical counterpart (i.e. were not a part of the low-z catalog curated by \cite{Wing2011}), indicating they were likely at high-z, or that were previously found to be associated with a $z>0.7$ luminous quasar. From their sample of IR-detected host galaxies, \cite{PM2013} form the high-z COBRA catalog. They determine photometric redshifts by modeling the infrared colors, with priors defined by assuming the radio host lies at $z>0.7$, and identified candidate galaxy clusters through 2$\sigma$ excesses in the galaxy number counts, compared to a background field, surrounding the bent radio AGN. \cite{GM2019} use further follow-up optical photometry from the Lowell Discovery Telescope (LDT; previously known as the Discovery Channel Telescope or DCT) in the SDSS \emph{r-} and \emph{i-}band filters for the high-z COBRA sources to identify overdensities of red galaxies surrounding the high-z COBRA targets. Clusters were further confirmed using a red sequence selection using the optical and IR data. Thus, a small subset of the high-z COBRA catalog have detections in both the IR, with Spitzer, and deep optical imaging.

The focus of this paper is one of the quasar-associated high-z COBRA cluster candidates, \targ. It was selected for spectroscopic follow-up because it was one of the highest-redshift (with a confirmed spectroscopic redshift of $z_{spec} = 1.8157$ \citep[SDSS-DR16 Quasar Catalog;][]{Lyke2020} for the quasar host) COBRA cluster candidates that showed a significant overdensity of IR-selected and optically-red galaxies surrounding the bent radio AGN \citep{PM2017, GM2019}. 

Here we present follow-up HST/WFC3 grism observations of \targ. From the grism spectra, we obtain redshifts for hundreds of sources in the field surrounding \targ, allowing us to spectroscopically identify members of this candidate cluster. Additional follow-up observations of \targ\ were taken with the Chandra X-ray Observatory with the aim of identifying ICM emission which would allow further confirmation of the high-z cluster candidate. The $\sim$100ks X-ray observations will be the focus of a separate, follow-up paper (in prep.).

A summary of our HST observations, including additional optical and IR photometry, is presented in Section \ref{sec:obs}. Sections \ref{sec:grizreduc} \& \ref{sec:grizfit} detail our reduction and redshift fitting of the grism spectra. Section \ref{sec:results} summarizes the galaxy cluster member selection technique and discusses individual galaxy properties. In Section \ref{sec:concl}, we summarize our findings and discuss our conclusions.

Throughout this paper, we adopt a standard $\Lambda$CDM cosmology with $H_0 = 70$ km s$^{-1}$ Mpc$^{-1}$, $\Omega_M = 0.3$, and $\Omega_{\Lambda} = 0.7$. At the redshift we find for \targ\ ($z=1.81$), the luminosity distance is $D_L = 13.7$ Gpc, the angular diameter distance is $D_A = 1.74$ Gpc, and $1\arcsec = 8.44$ kpc. Uncertainties reported here are 1$\sigma$ confidence intervals unless otherwise noted. All magnitudes are reported in the AB Magnitude system unless otherwise noted.

\section{Observations\label{sec:obs}}
\subsection{HST Grism Observations}

\begin{figure*}
\plotone{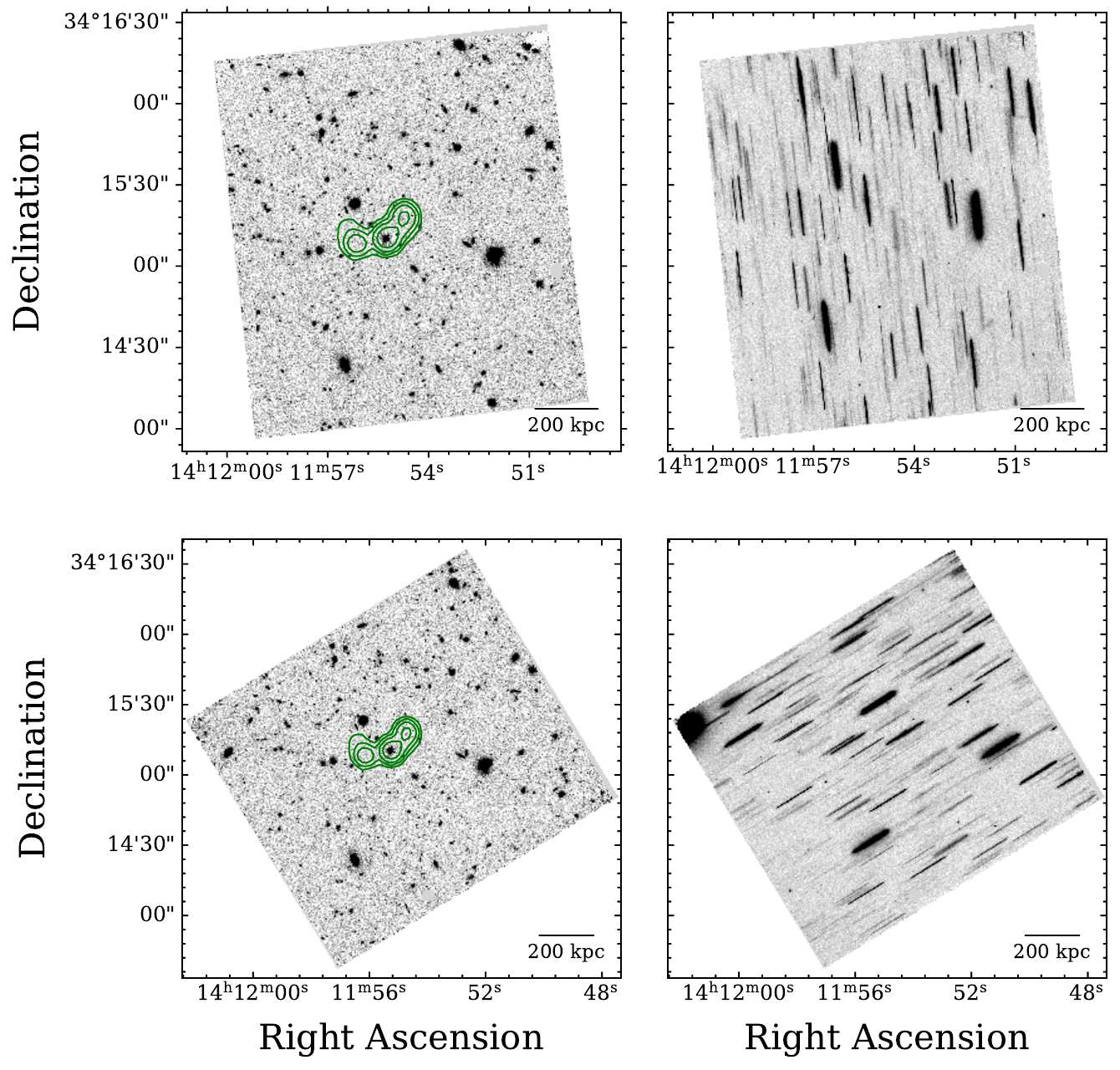}
\caption{HST F140W direct images (\emph{left}) and G141 grism spectra (\emph{right}) of \targ\ at each PA orientation. All images have pixel scales of 0.13\arcsec/px and measure $\sim$3\arcmin$\times$3\arcmin. Overlaid on the F140W images are the VLA FIRST 20 cm radio contours (green) of the COBRA source. The scale bar is defined at the redshift of the target quasar.}\label{fig:hstimgs}
\end{figure*}

After the identification of the potential clusters in the COBRA survey (\cite{PM2017}; hereafter \citetalias{PM2017}), follow-up observations were performed using the Wide Field Camera-3 (WFC3) onboard the Hubble Space Telescope (HST) for \targ\ (GO proposal 15994; PI Blanton). Observations were taken in two visits using nearly orthogonal orientations to allow for the deblending of neighboring spectra. For each visit, we obtained 2 ks of G141 grism exposures with an accompanying 0.5 ks of F140W direct imaging taken just before the grism observations. Thus, the total exposure times of our observations are 4 ks in G141 and 1 ks in F140W. Figure \ref{fig:hstimgs} shows the G141 grism image with corresponding F140W direct image at each of the observed position angle (PA) orientations (\verb|PA_V3| = 51$^{\circ}$ and 346$^{\circ}$). Each image has a pixel scale of 0.13\arcsec/pix, covering a 3\arcmin$\times$3\arcmin\ area.

The G141 grism covers a wavelength range of 1075–1700 nm, with a dispersion of 4.65 nm/pixel and a resolution of $R\sim130$. At a redshift of $z=1.81$ (the redshift of the quasar that hosts the bent, double-lobed radio source), the G141 grism allows for the identification of strong spectral features such as H$\beta$ and [OIII]. Our shallower observations limit us only to the detection of star-forming galaxies and we cannot confirm any quiescent galaxies, which lack the requisite emission lines.

\subsection{Additional Photometry \label{sec:addphot}}

\begin{figure*}
    \centering
    \plotone{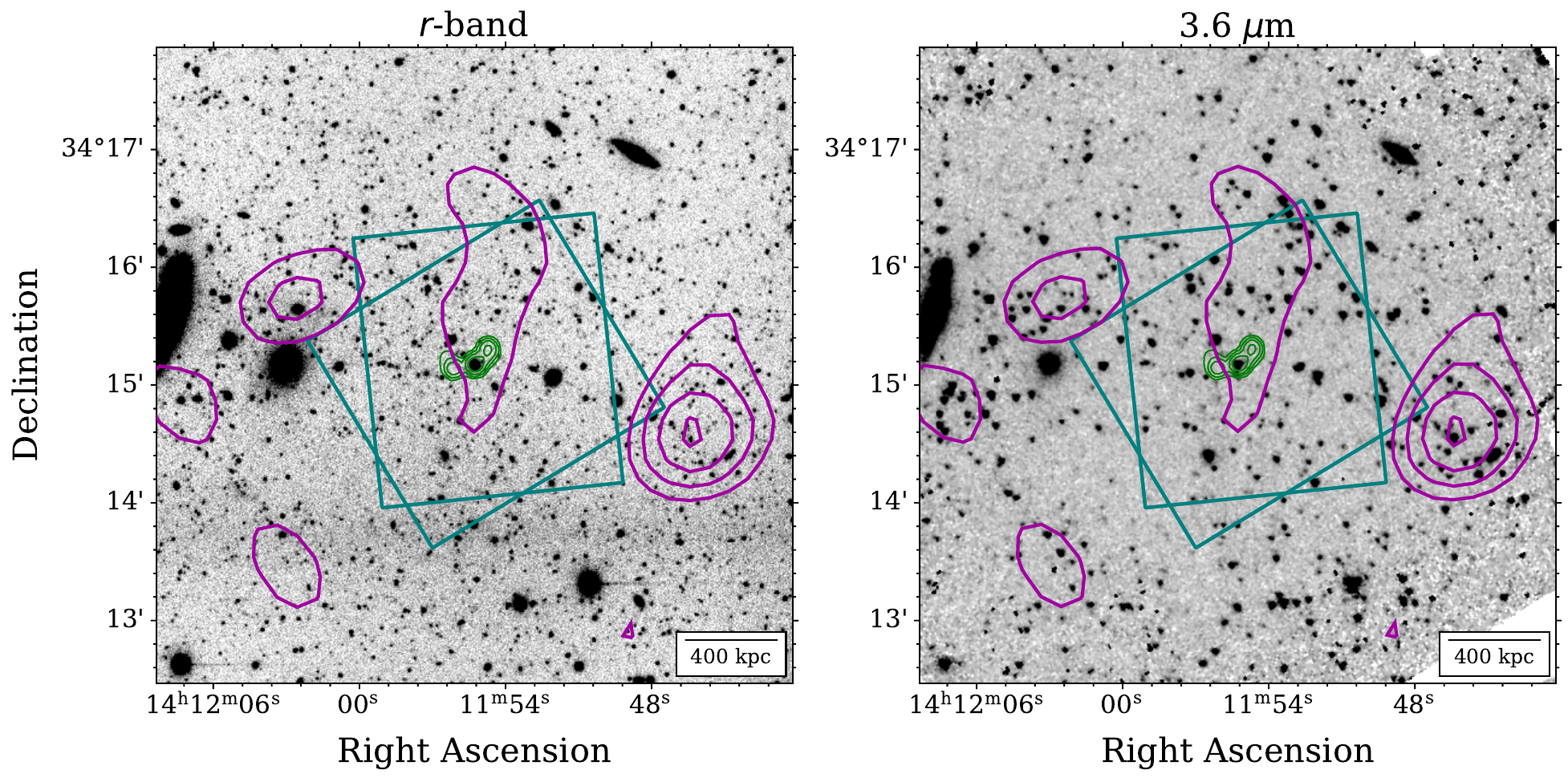}
    \caption{LDT $r-$band (\emph{left}; \citetalias{GM2019}) and Spitzer 3.6 \micron\ (\emph{right}; \citetalias{PM2017}) mosaics of \targ. Both panels display a 5.4\arcmin$\times$5.4\arcmin\ area, centered on the target quasar. Green contours show the VLA FIRST 20 cm radio contours of the bent, double-lobed radio source. Magenta contours show the projected density of red (high-z candidate) galaxies as measured by \citetalias{GM2019}. The footprints of our HST observations are outlined in cyan. The scale bar is defined at the SDSS spectroscopic redshift of the bent, double-lobe radio source host quasar ($z=1.81$).}
    \label{fig:ldt_spz}
\end{figure*}

\targ\ was previously observed in the mid-infrared, using Spitzer/IRAC at 3.6 \micron\ and 4.5 \micron\ \citepalias{PM2017} and in the optical, using the LDT in the SDSS \emph{i-} and \emph{r-}band filters (\cite{GM2019}; hereafter \citetalias{GM2019}). Figure \ref{fig:ldt_spz} shows the LDT r-band and Spitzer 3.6 \micron\ images, covering a 5\farcm4 $\times$5\farcm4\ area, centered on the quasar source, which signaled \targ\ as a potential high-z cluster. Overlaid on both panels in Fig.\ \ref{fig:ldt_spz} are the footprints (cyan) of the HST observations presented here, contours of VLA FIRST 20 cm emission (green) of the bent, double-lobed radio source associated with the target quasar, and the projected density of red (high-z candidate) galaxies (magenta) as measured by \citetalias{GM2019}.

\citetalias{PM2017} used the Spitzer observations to search for galaxy overdensities in the high-z COBRA sample. For \targ, they find an overdensity of 17 sources within a 1\arcmin\ radius, corresponding to a 2.6$\sigma$ significance. Within 2\arcmin, this detection increases to 3$\sigma$ with an excess of 44 sources. The 3.6 \micron\ and 4.5 \micron\ photometry are included in the redshift fitting (see \S \ref{sec:grizfit}) when available. Because the Spitzer/IRAC spatial resolution is only 2\arcsec, each object with recorded mid-infrared photometry is visually inspected to identify instances of blended sources. In these cases, the IRAC photometry is excluded from the fit to the SED. 

Continuing on \citetalias{PM2017}'s mid-IR work with the high-z COBRA sample, \citetalias{GM2019} used follow-up optical observations in conjunction with the Spitzer observations to identify red-sequence overdensities. For \targ, they identify 12 red-sequence members within 1\arcmin\ of the radio AGN. Due to the high-z nature of \targ, \citetalias{GM2019} were unable to measure the significance of the red sequence using the optical observations, and therefore relied entirely on the $(m_{3.6}-m_{4.5})_{AB} > -0.15$ color cut similar to that of \cite{Papovich2010}. However, while the optical photometry were not useful for characterizing the cluster's red sequence, we do include the measurements in the redshift fitting below, when available for a given source. 

Given the depth of the Spitzer observations, \citetalias{GM2019} use the relatively shallow magnitude limit of 21.4 for source detection in both 3.6 and 4.5 \micron\ bands, which we choose to mirror here. Additionally, we adopt their optical magnitude limits of $m_i \leq 24$ and $m_r \leq 25$, which corresponds to the limiting magnitudes of the LDT observations. 

\begin{figure*}
    \centering
    \plotone{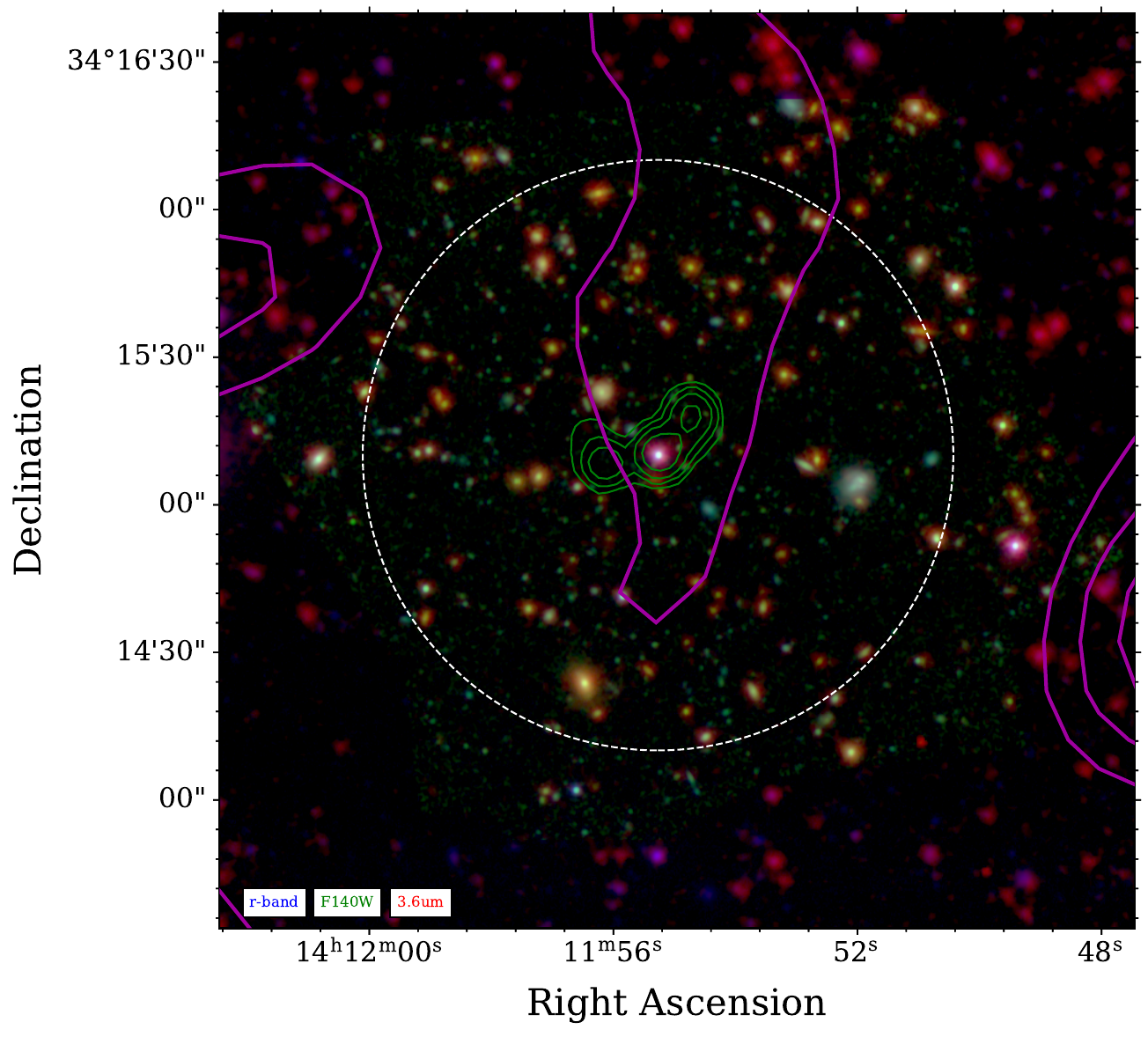}
    \caption{RGB mosaic of \targ\ composed of the Spitzer 3.6 \micron\ (red), HST F140W (green), and LDT r-band (blue) imaging. The F140W image has been lightly smoothed using a Gaussian kernel of radius $\sim0.7$\arcsec\ in order to better match the larger PSF of the IRAC imaging, however, the resolution of the three layers remains different. The green contours are the VLA FIRST 20 cm radio contours associated with the quasar source. Magenta contours show the projected density of red (high-z candidate) galaxies as measured by \citetalias{GM2019}. The white dashed circle measures 1\arcmin\ ($\sim 0.5$ Mpc at $z=1.81$) and is centered on the bent, double-lobed radio source.}
    \label{fig:rgb}
\end{figure*}

Figure \ref{fig:rgb} shows an RGB mosaic of \targ\ that combines the LDT r-band (blue; \citetalias{GM2019}), the HST F140W (green; this work), and the Spitzer 3.6 \micron\ (red; \citetalias{PM2017}) observations. The green contours are the VLA FIRST 20 cm emission of the bent radio AGN which tagged \targ\ as a potential high-z cluster candidate. The magenta contours show the projected density of red (high-z candidate) galaxies as measured by \citetalias{GM2019}. The white dashed circle measures 1\arcmin\ ($\sim$0.5 Mpc at $z=1.81$) in radius and is centered on the quasar source.

\section{Grism Reduction\label{sec:grizreduc}}

The HST G141 grism and F140W direct imaging observations were reduced using the Grism Redshift and Line analysis software, \textsc{Grizli}\footnote{\href {https://github.com/gbrammer/grizli/}{https://github.com/gbrammer/grizli/}} \citep[v1.3.2;][]{Brammer2021}. We refer the reader to the official documentation for a detailed description of the \griz\ reduction process. Here, we provide a short summary of the processes relevant to this paper.

Starting from the raw flt files obtained from MAST\footnote{Mikulski Archive for Space Telescopes: \href {https://archive.stsci.edu}{https://archive.stsci.edu}}, \griz\ is used to build associations between the grism exposures and the corresponding direct images. All the necessary preprocessing steps for both the F140W and G141 exposures, including flat-fielding, cosmic-ray rejection, background subtraction\footnote{Using the master sky backgrounds of \cite{Brammer2015}}, and fine astrometric alignment\footnote{Individual exposures were aligned to reference sources in the GAIA \citep{Collaboration2016} Early Data Release 3 (EDR3) \citep{Gaia_Collaboration2021}} are performed using \griz. After the preprocessing and aligning is completed, \griz\ is used to create drizzled mosaics\footnote{Utilizing \textsc{AstroDrizzle} \citep{Gonzaga2012}} of the F140W direct imaging, with a pixel scale of 0.13\arcsec/px. A photometric catalog is created from the drizzled direct image mosaic using \textsc{sep}\footnote{A python implementation of Source Extractor \citep{Bertin1996}.} \citep{Barbary2016} for source detection and aperture photometry. The positions of sources detected in the field are then used to identify the spectral traces for each object and when modeling contamination from neighboring sources.

A full field contamination model is created for each grism exposure using \griz\ through an iterative process: (1) A first-pass model of the full detector field of view is created assuming simple linear continua for all objects in the field; 
(2) The model is refined further through third-order polynomial fits directly to the observed spectra after subtraction of the first-pass contamination model. The subtraction of contamination from neighboring/overlapping sources is aided by the orthogonal orientations of the observed PAs.

For each source identified in the F140W reference image, a 2D spectral cutout is extracted from each individual G141 exposure, resulting in a maximum of 8 grism extractions per object. The final contamination model is subtracted from the individual 2D grism exposures to remove contamination from neighboring sources. 

After the reduction and extraction is complete, for each source we have data products which include 2D grism cutouts of the individual G141 exposures, a stacked 2D grism spectrum combining both PAs, the contamination model in the region of the source, and an F140W direct image thumbnail. 

\section{Grism Redshift Fitting\label{sec:grizfit}}

\griz\ is used to determine the redshifts of all sources detected in the reference image by template fitting the contamination-subtracted grism spectra. Spitzer and LDT photometry, when available, are included (as fluxes which were converted from the reported AB magnitudes) as input to the \griz\ fitting routine as long as $m_{3.6}, m_{4.5} < 21.4$, $m_i<24$, and $m_r<25$, with the limits following the convention of \citetalias{GM2019}.

Fitting is performed in a two-step iterative process within \griz. A first pass fit is performed over a coarse (dz=0.004) redshift grid from $z=0-4$ using the default suite of Flexible Stellar Population Synthesis (FSPS; \cite{Conroy2009, Conroy2010}) templates and emission line complexes provided by \griz. These FSPS templates are designed to represent model galaxies with a wide range of properties \citep{Brammer2008}. A second pass fitting is performed using a finer (dz=0.0004) redshift grid defined around the highest peak in the probability distribution function from the first-pass fit results, using the same suite of FSPS templates but now allowing fits of individual emission lines. This also allows for the determination of line fluxes of the fitted emission lines. 

The resulting data products include the 1D contamination subtracted source spectrum, full redshift solution on both the coarse and fine grids, the covariance matrix and $\chi^2(z)$ of the template SED fit, the best-fit template SED with emission lines, line fluxes with errors and equivalent widths of individual emission lines fit, and the stellar population synthesis (SPS) parameters calculated from the FSPS templates. We refer the reader to \cite{Abramson2020} for an in-depth description of all the data products of the \griz\ redshift fitting process.

\section{Results \& Discussion \label{sec:results}}

The source catalog that was generated using \textsc{SEP} during the reduction process (see \S \ref{sec:grizreduc}) identified 450 sources in the HST field of view. In our analysis of source counts as a function of magnitude, we find that the number of detected sources begins to drop sharply at magnitudes fainter than 25, corresponding to our magnitude limit. Thus, we consider only those sources with magnitudes brighter than $m_{F140W} \leq 25$, leaving 291 grism objects to fit. Each source is fit individually, with visual inspection of the extracted spectra and the resulting fit to assess reliability. Of these, we obtain robust grism redshifts for 147 sources. Table \ref{tab:allfits} lists the coordinates, F140W AB magnitudes, and best-fit grism redshifts, with 1-$\sigma$ errors, for a few of these sources. For the full catalog of sources, we refer the reader to the full version of this table, available as an online supplement to this paper.

\begin{figure}
    \centering
    \includegraphics[width=\linewidth]{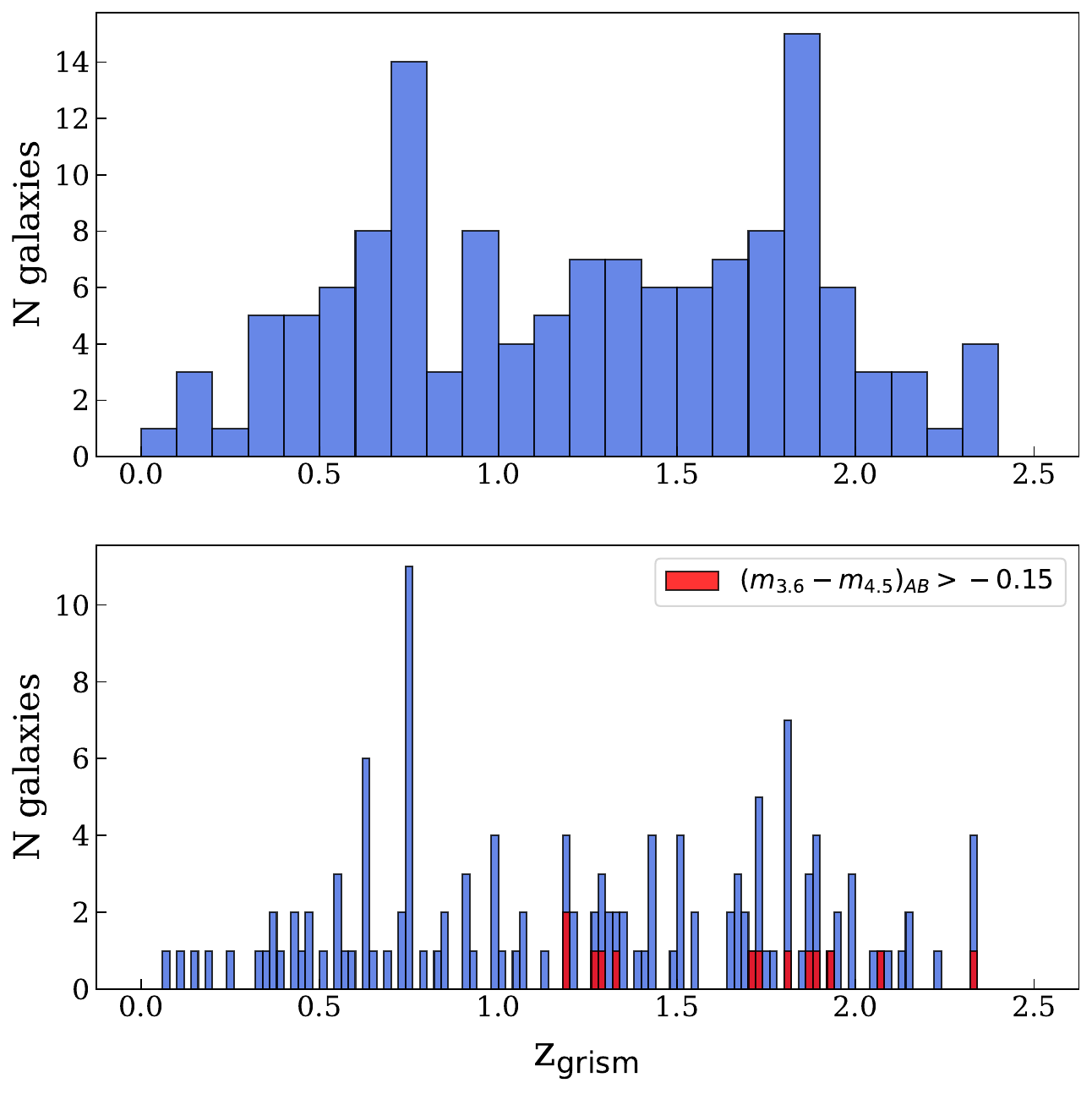}
    \caption{Distribution of HST grism redshifts in \targ\ at $z_{grism}\leq 2.5$ with bin widths corresponding to 0.1 (top panel) and 0.02 (bottom panel) in redshift. Sources that satisfy the $(m_{3.6}-m_{4.5})_{AB} > -0.15$ color cut of \citetalias{GM2019} are indicated in the bottom panel in red. We find strong peaks at $z\sim 0.75$, suggesting a possible foreground structure, and at $z\sim 1.8$, which corresponds to the redshift of the quasar hosting the bent, double-lobed radio source.}
    \label{fig:zhist}
\end{figure}

Figure \ref{fig:zhist} shows the grism redshift distribution for sources in \targ\ in the range $0\leq z_{grism}\leq 2.5$.  There are two notable peaks in redshift. The first peak at $z\sim 0.7$ could indicate a potential foreground structure. We focus in on the second peak at $z\sim 1.8$, the redshift of the quasar source in \targ, for the rest of the analysis presented here. The bottom panel of Fig.\ \ref{fig:zhist} shows the redshift distribution with bin widths of 0.02 in redshift. This is discussed further in \S \ref{sec:cands}. 

\targ\ was initially identified as a candidate cluster in both \citetalias{PM2017} \& \citetalias{GM2019} through the detection of an overdensity of red sources. In addition, \citetalias{GM2019} identify 12 sources within 1\arcmin\ of the quasar which hosts the bent, double-lobed radio source which satisfy the $(m_{3.6}-m_{4.5})_{AB} > -0.15$ color-cut, indicating the presence of a population of galaxies at redshifts $>1.3$. We are able to obtain reliable redshifts for 11 of these sources. Of these, nine sources are fit at $z>1.3$, consistent with their color selection, with the two outliers both lying at $z=1.19$. Note that \citetalias{GM2019} adopted the shallow limits of $m_{3.6}, m_{4.5} \leq 21.4$ when considering cluster members based on color so there could be a larger number of passive red galaxies that we are missing. Therefore, full cluster membership could be greater than what we are able to spectroscopically confirm here. 

\begin{figure*}
\centering
\begin{tabular}{cccc}
  \includegraphics[width=0.33\textwidth]{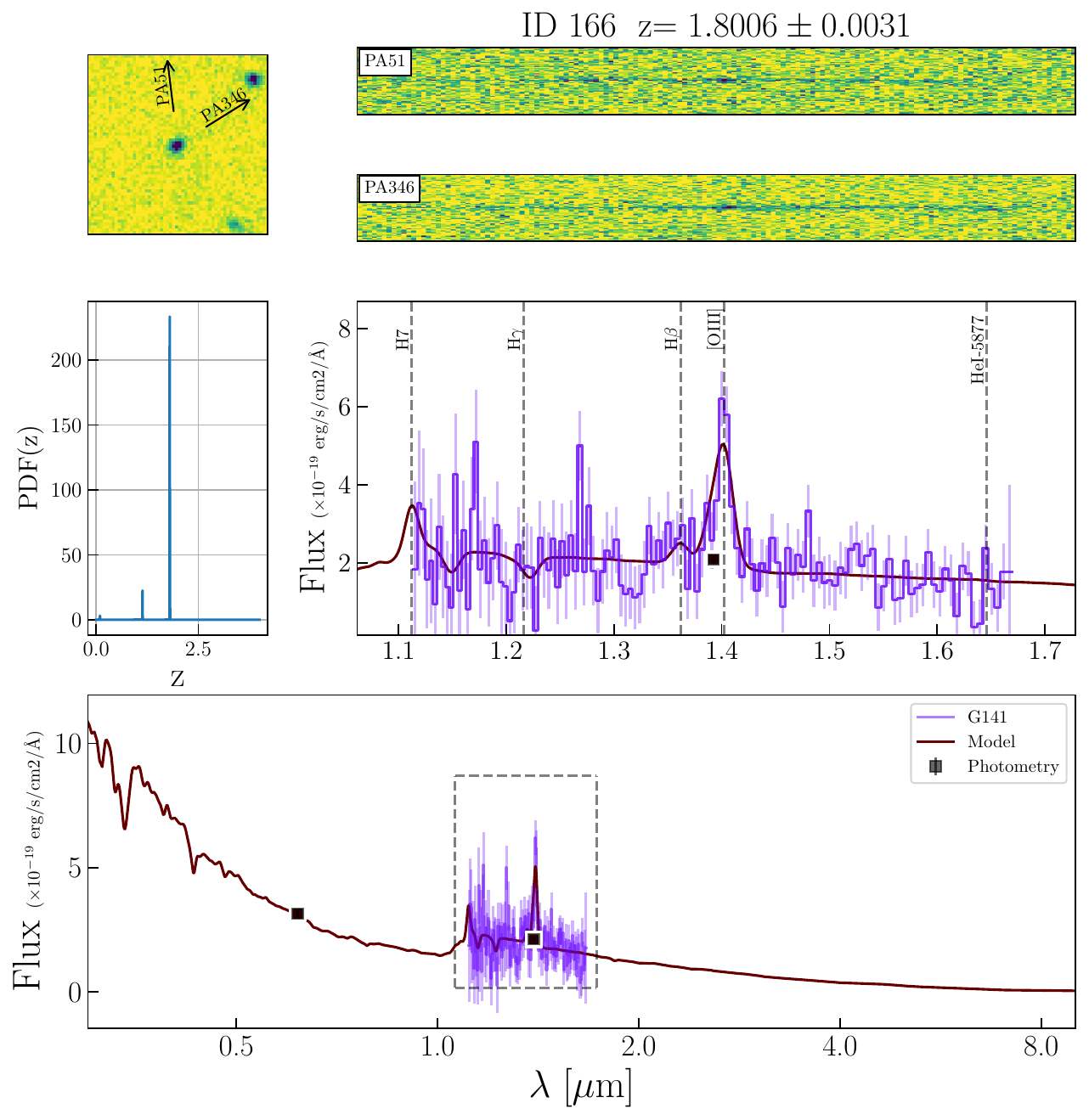} & &   \includegraphics[width=0.33\textwidth]{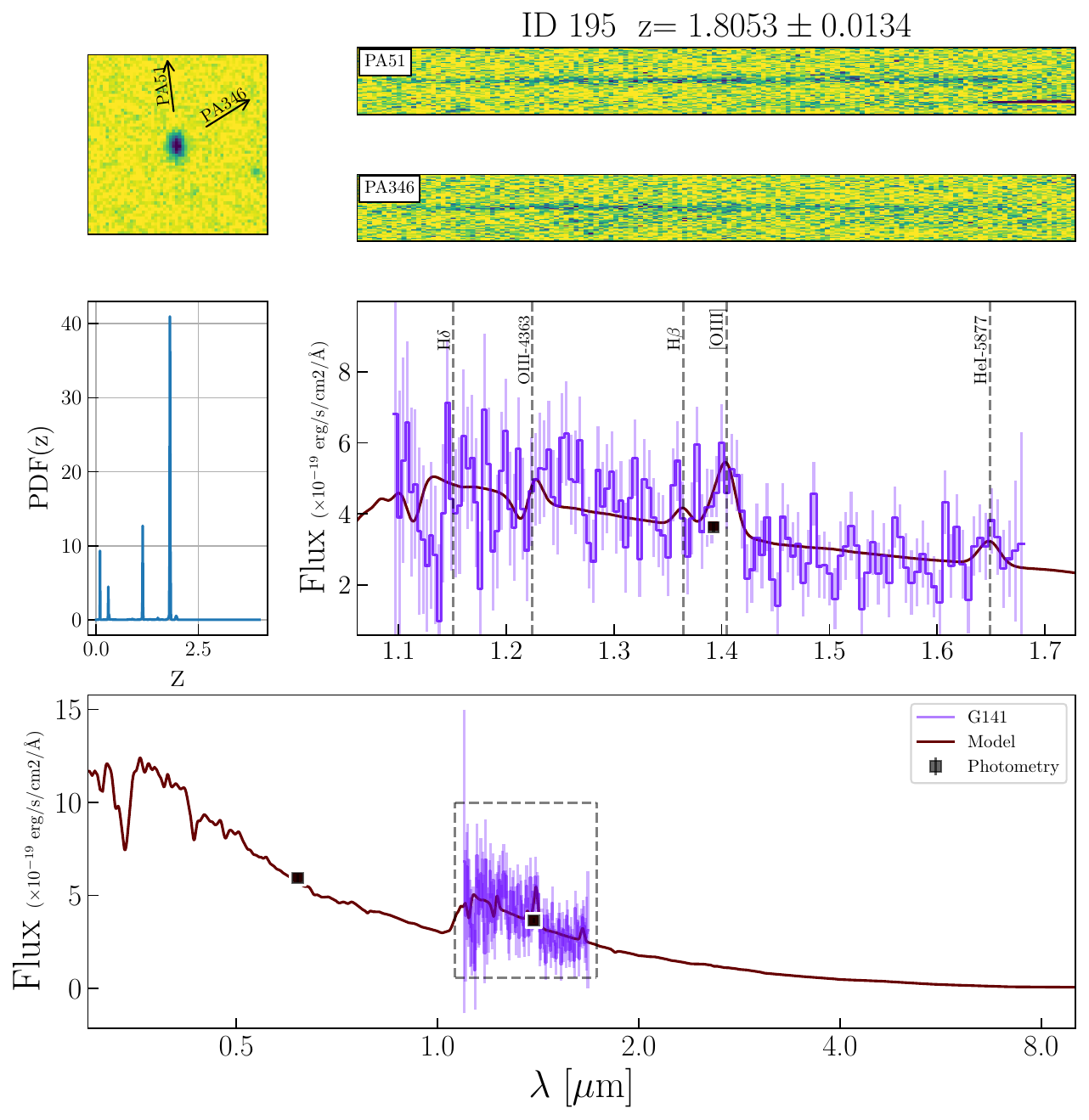} & 
  \includegraphics[width=0.33\textwidth]{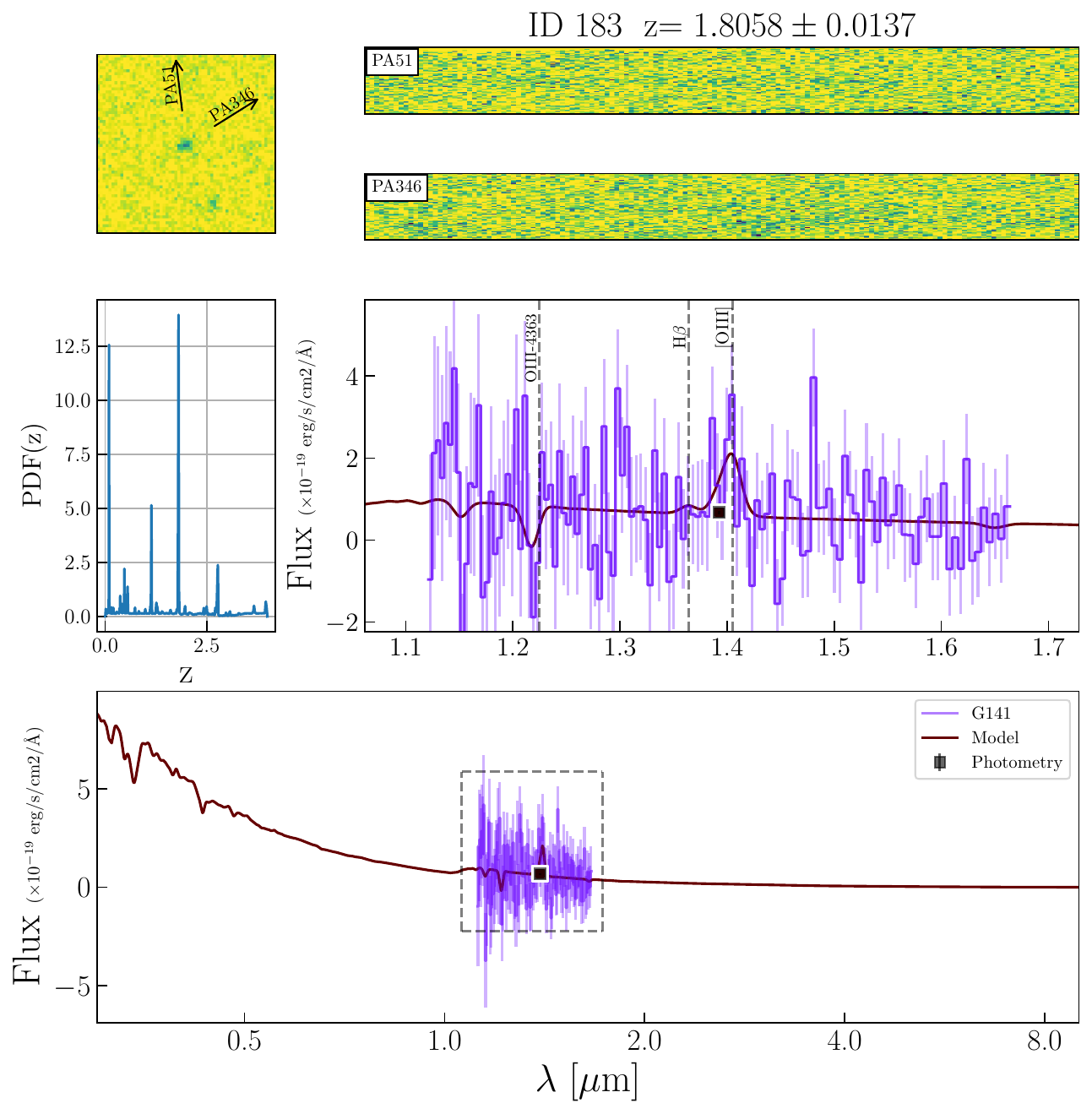} \\
 \includegraphics[width=0.33\textwidth]{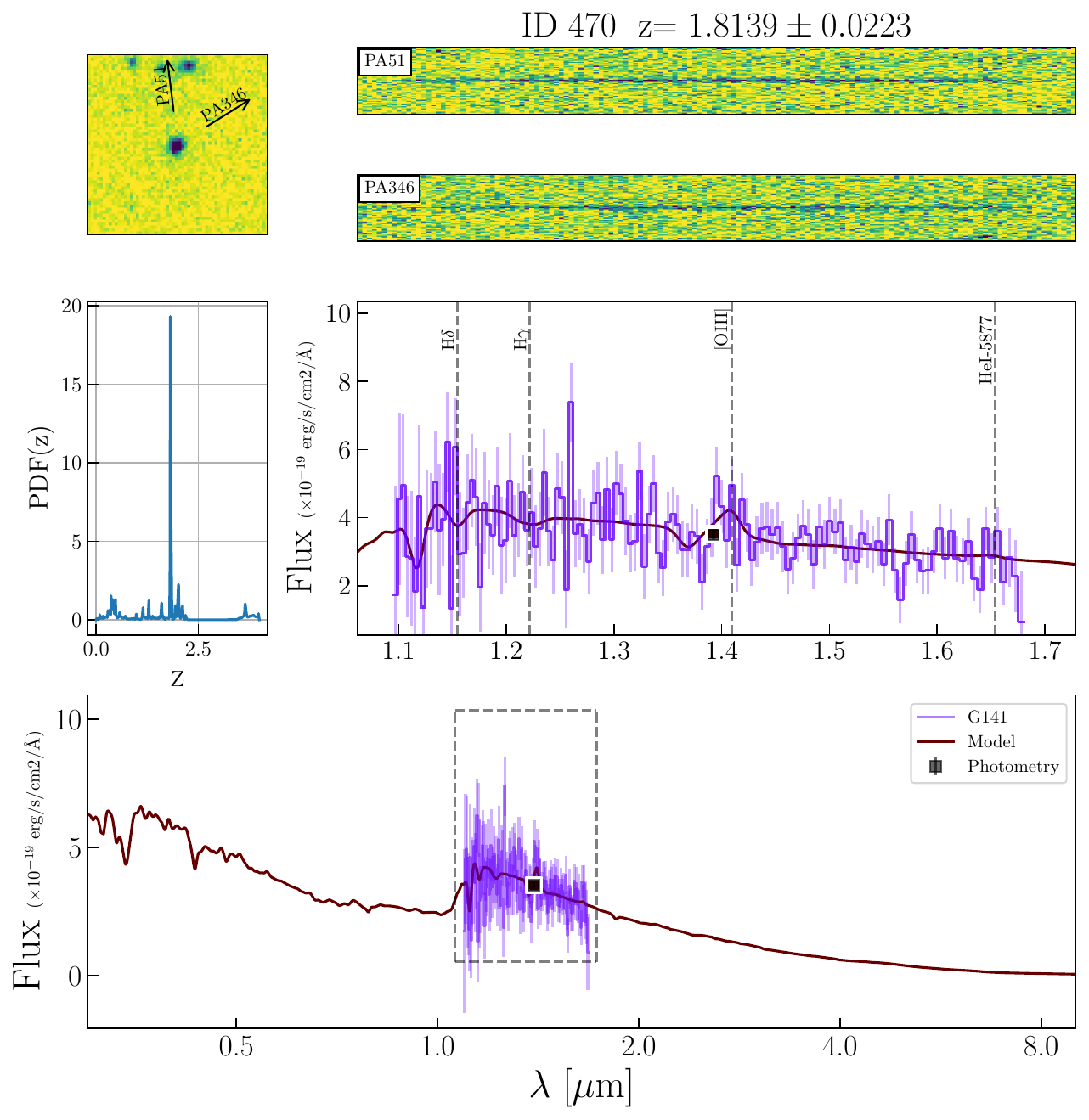} & &   \includegraphics[width=0.33\textwidth]{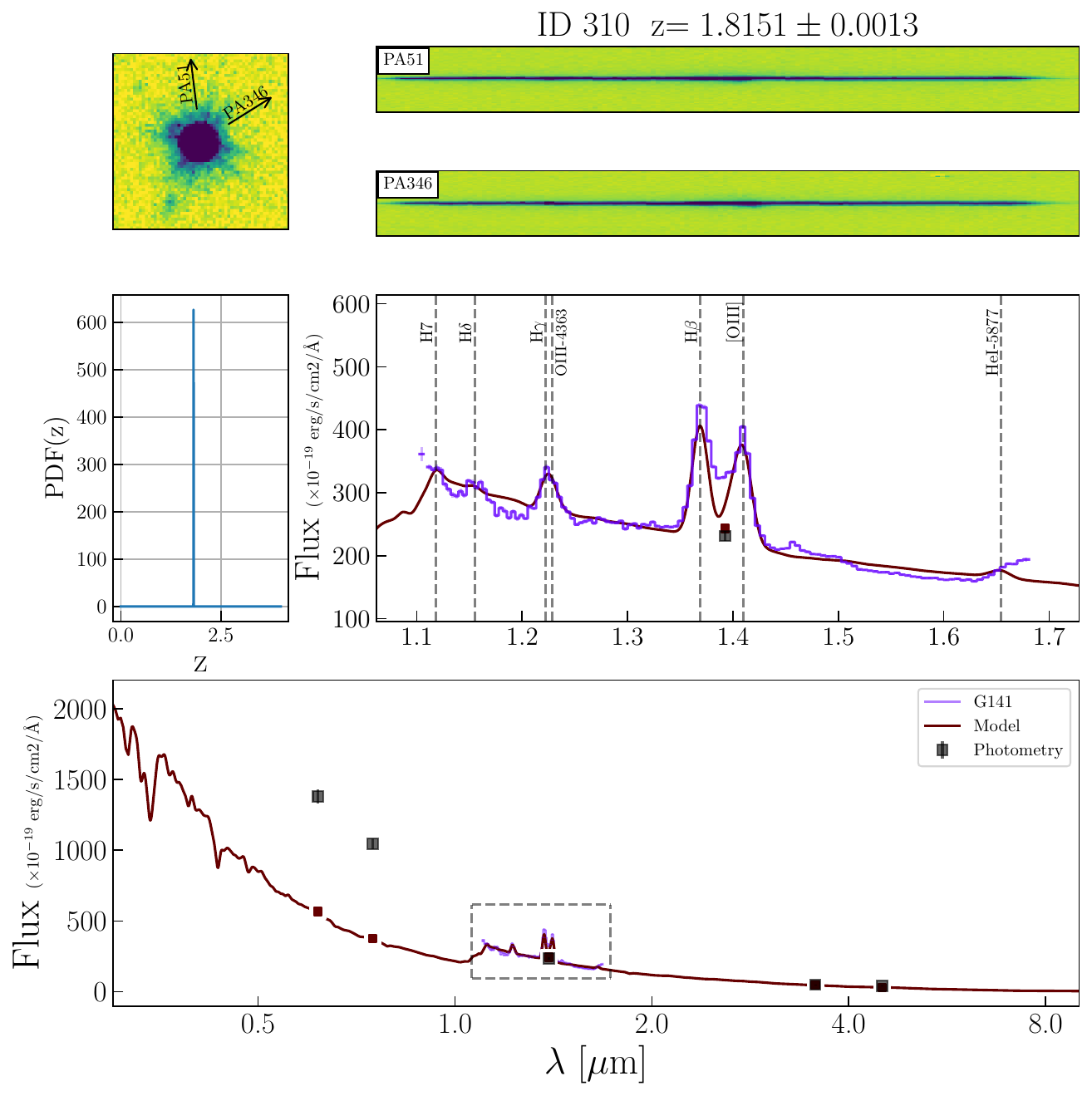} & 
 \includegraphics[width=0.33\textwidth]{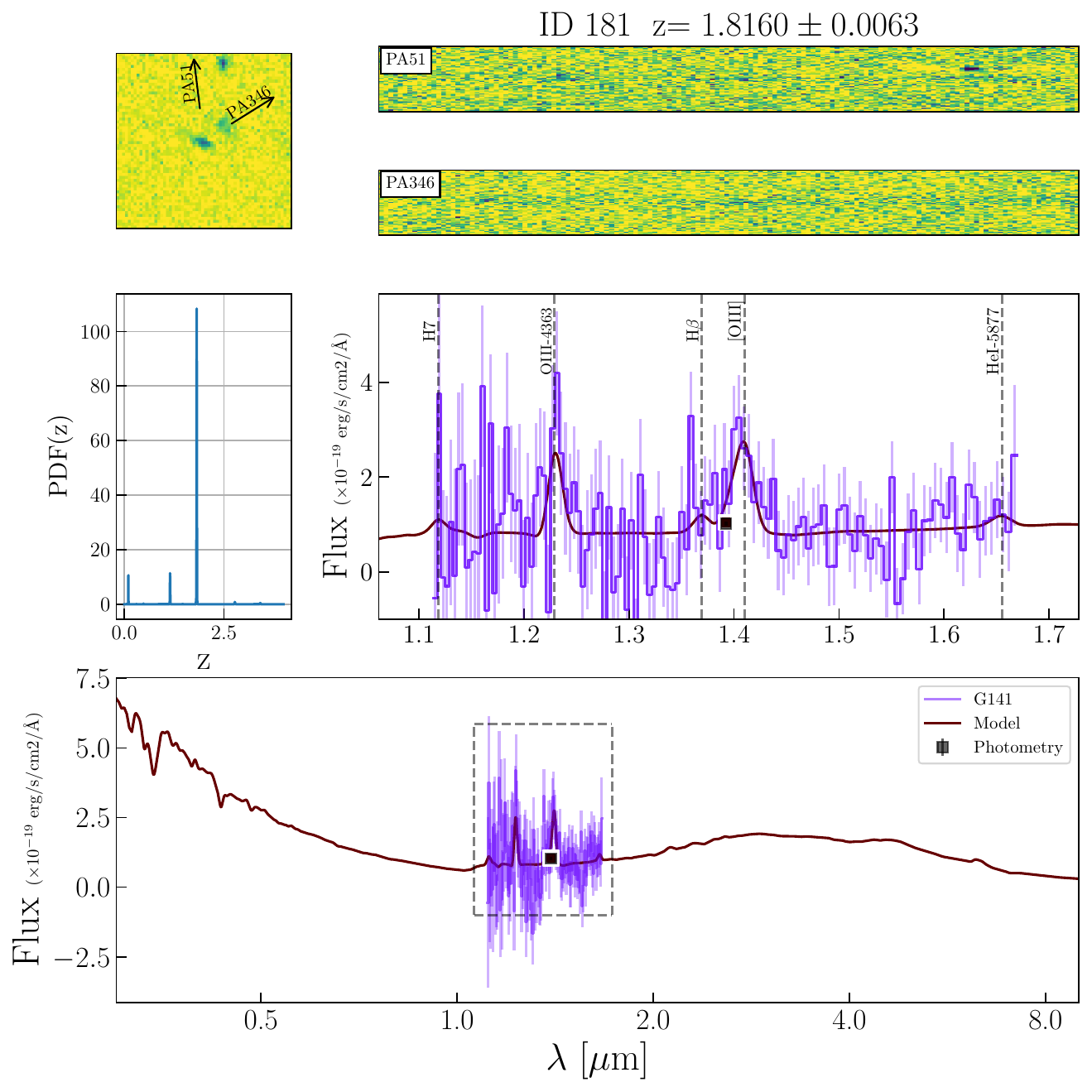}\\
 \includegraphics[width=0.33\textwidth]{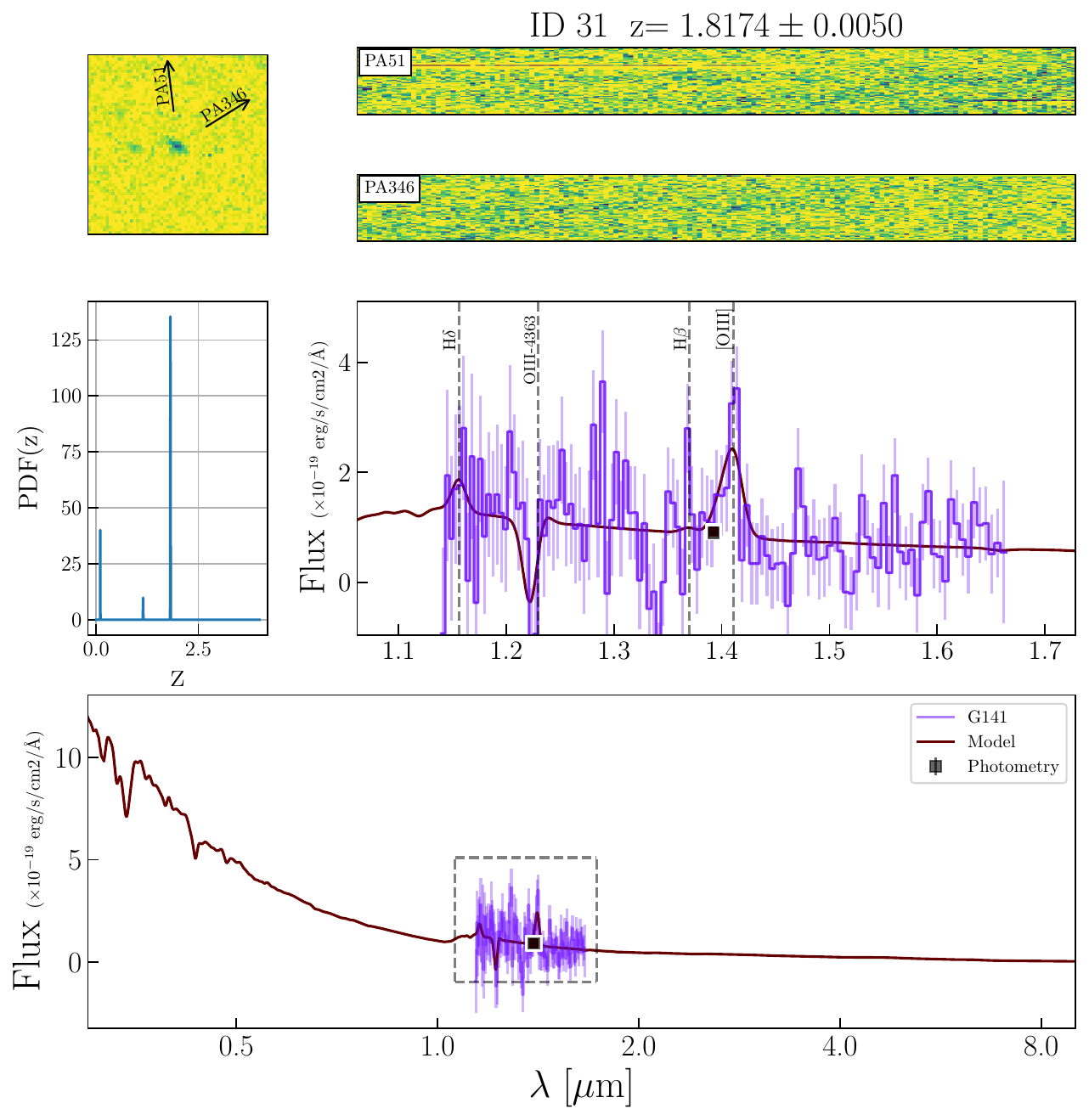} & {\vrule width 1pt} &   
 \includegraphics[width=0.33\textwidth]{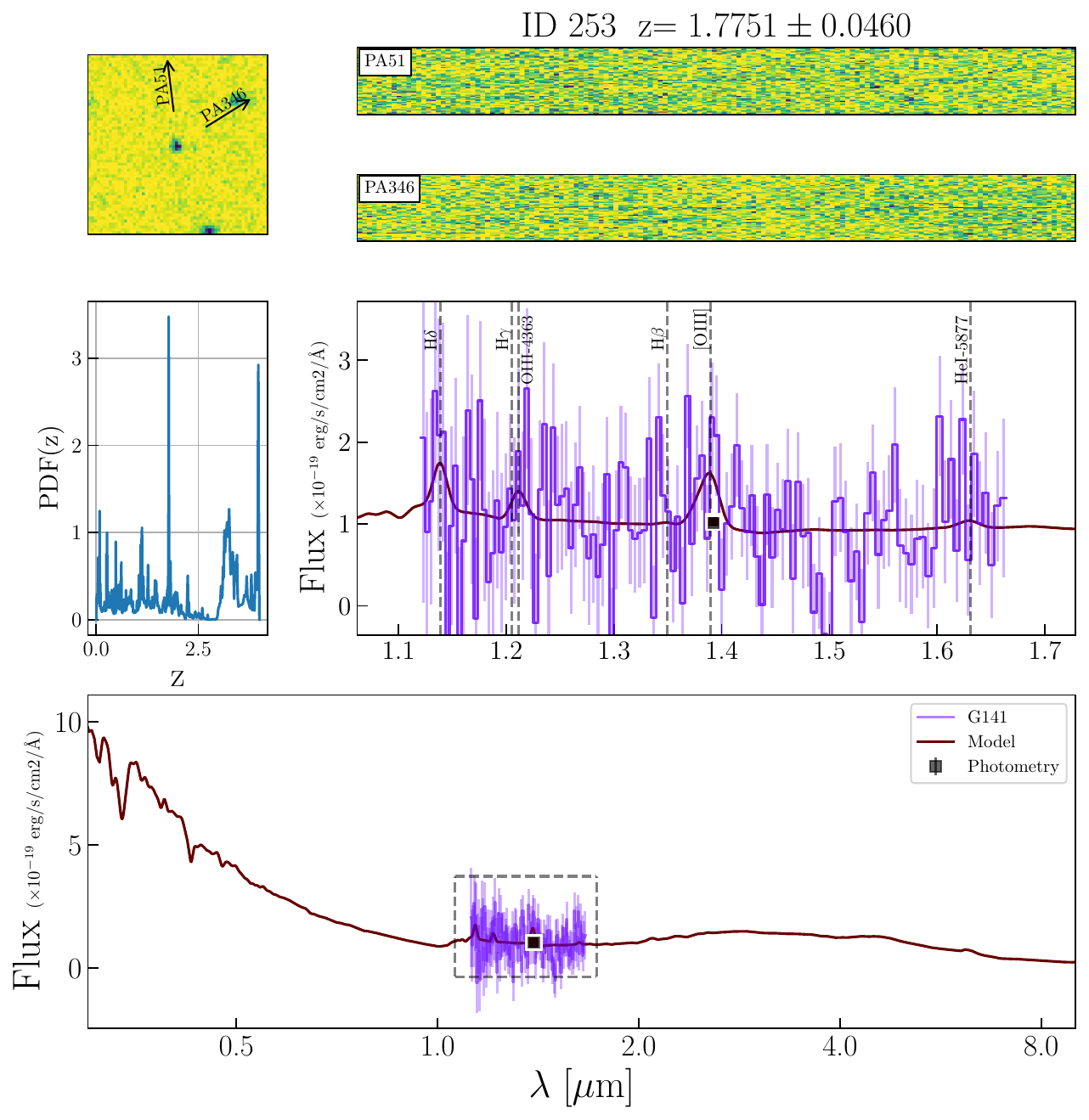} &
 \includegraphics[width=0.33\textwidth]{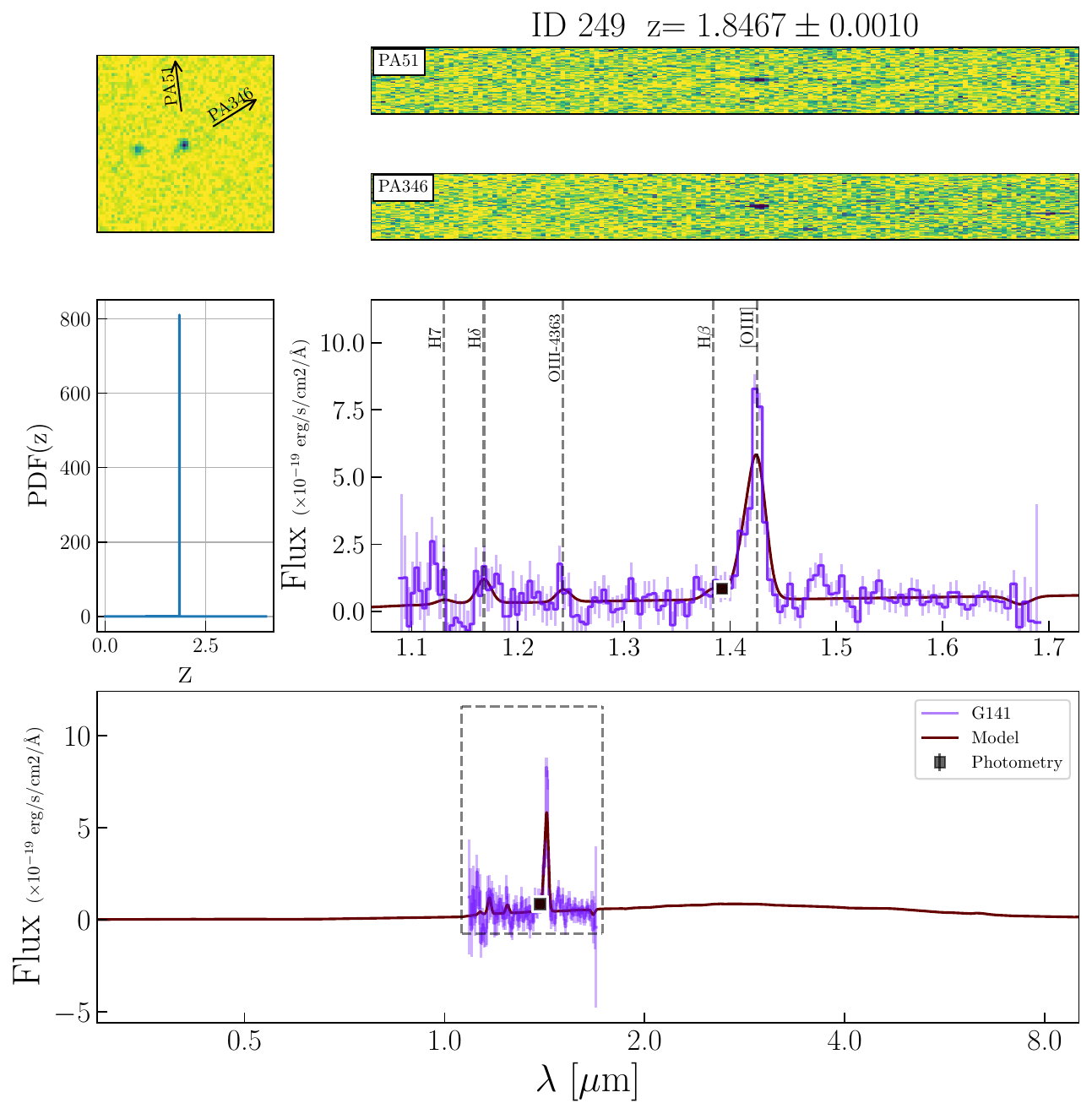}\\
\end{tabular}
\caption{HST G141 grism spectra and redshift fits for members of \targ\ (see \S\ref{sec:cands} for selection process), arranged in order of increasing redshift, followed by the two possible infalling members (separated by the vertical black line). For each source we show: a 6\arcsec$\times$6\arcsec\ F140W direct image thumbnail (top left panel), the G141 grism cut-out (top right panels), the redshift probability distribution function (PDF; center left), the 1D grism spectrum with \griz\ best-fit model over the G141 wavelength range (center right), and the full best-fit SED template model (bottom panel) with additional IR and/or optical photometry included when present. The arrows on the F140W thumbnail denote the PA direction of each G141 cut-out. \label{fig:candfits}}
\end{figure*}

\subsection{Identifying Candidate Cluster Members\label{sec:cands}}

In order to confirm \targ\ as a bona-fide high-z cluster, we first establish selection criteria for member galaxies. We adopt similar criteria to those of \cite{Noirot2018}, which is a modified version of the \citet{Eisenhardt2008} criteria. \cite{Noirot2018} define a confirmed cluster as a structure having at least five galaxies within a physical radius of 0.5 Mpc whose spectroscopic redshifts lie within $\pm$ 2,000 km/s. 

We find a best-fit grism redshift of $z_{grism} = 1.8151\pm0.0013$ for the quasar that hosts the bent, double-lobed radio source which flagged \targ\ as a cluster candidate in the COBRA survey. We identify six additional sources within $\pm$ 2,000 km/s (corresponding to a redshift range of $1.7963 \leq z_{grism}\leq  1.8339$). Thus, we have identified seven potential members, satisfying the cluster criteria of \cite{Noirot2018}. There is a mean redshift of $\langle z\rangle=1.8106$ with a deviation of $0.006$ between redshift in these seven members which span the range $z=1.8006-1.8175$.

We re-evaluate the redshift distribution (top panel of Fig.\ \ref{fig:zhist}), but with bins centered on the redshift range of our seven cluster members ($z=1.80-1.82$) and a binning of 0.02 in redshift. This updated histogram is shown in the bottom panel of Fig.\ \ref{fig:zhist}, and also highlights sources that satisfy the $(m_{3.6}-m_{4.5})_{AB} > -0.15$ color cut of \citetalias{GM2019}. With this updated analysis, we find that the two prominent peaks near $z\simeq 0.7$ and $z\simeq 1.8$ are again the two most notable peaks and are now resolved into eleven sources at $z=0.74-0.76$ and our seven cluster members at $z=1.80-1.82$, the latter of which is significant at $5\sigma$ above the background. Excluding the peaks near 0.7 and 1.8, we find a mean number of galaxies of 0.96$\pm$1.26 per 0.02 redshift bin spanning redshifts 0 to 2. We take this mean as the ``background'' distribution of galaxies.

Following the methods of \cite{Blanton2000}, we calculate the line-of-sight velocity dispersion for \targ\ candidate members of $\sigma_{\parallel} = 701^{+347}_{-138}$ km/s, where the errors are the 68\% confidence uncertainties calculated following the convention of \cite{Danese1980}. 

In order to ensure that we are not including unrelated field galaxies as potential cluster members, we ``clean'' our data using the 3$\sigma_{\parallel}$ clipping methods of \cite{Yahil1977} and \cite{Blanton2000}. We begin by removing the galaxy with the largest velocity offset from the cluster mean, which in the case of \targ\ is object 166, and recalculate $\sigma_{\parallel}$. The velocity offset of source 166 is recalculated from the new cluster mean and is found to be 2.2$\sigma_{\parallel}$ which satisfies the 3$\sigma_{\parallel}$ cut. Therefore, we conclude that no unrelated field galaxies are included in the candidate members identified in \targ. 

Table \ref{tab: candmembers} lists the coordinates (as measured from the F140W images) and photometry from LDT, HST, and Spitzer, for the seven spectroscopically identified cluster members, as well as the two additional possible infalling members. For sources that were not detected in the LDT and Spitzer imaging, we report the limiting magnitude for the given filter which represents the upper limit for that source in that filter. The best-fit grism redshift and calculated velocity offsets for the sources are listed in Table \ref{tab:fitinfo}. Figure \ref{fig:candfits} shows the 1D and 2D grism spectra along with the best-fit \griz\ model and redshift for the spectroscopically identified candidate members of \targ\ as well as the two additional possible infalling members. For each source, we show a 6\arcsec $\times$ 6\arcsec\ F140W direct image thumbnail (top left panel), the 2D grism cutouts at each PA (top right panels), the redshift probability distribution (PDF; center left panel) from the first pass \griz\ redshift fitting process, the 1D grism spectrum with \griz\ best-fit model over the G141 wavelength range (center right panel) and the full best-fit SED model (bottom panel) with additional photometry included if available. The quasar which signaled \targ\ as a potential galaxy cluster corresponds to object ID 310 in Table \ref{tab: candmembers} and Figs.\ \ref{fig:candfits}.

\subsubsection{Main $z=1.81$ Cluster Radius \& Mass \label{sec:clustermass}}

We estimate the $R_{200}$ and $M_{200}$, the mass and radius inside of which the density is equal to 200 times the critical density of the universe, i.e. $\rho(R_{200}) = 200\rho_c = 200 \frac{3 H^2(z)}{8\pi G}$ where $H(z) = H_0 \sqrt{\Omega_M(1+z)^3 + \Omega_\Lambda}$ and $G$ is the gravitational constant. We calculate a virial radius as \citep{Finn2005, Jaffe2013}:
\begin{equation}
    R_{200} = \frac{\sqrt{3}\sigma_{\parallel}}{10 H(z)}\quad \text{ Mpc}
\end{equation}
The virial mass ($M_{200} = \frac{4}{3}\pi R^3_{200} \rho(R_{200})$) is then 
\begin{equation}
    M_{200} = \frac{100 H^2(z)}{G} R^3_{200}\quad 
\end{equation}
Using the velocity dispersion we found for the \targ\ cluster, $\sigma_{\parallel} = $ \fe{701}{347}{138}, we estimate $R_{200} = $ \fe{0.64}{0.32}{0.13} Mpc --- which, at $z_{cl} = 1.8106$, is just about the size of our HST observation's FoV --- and $M_{200} = $ \fe{2.2}{3.3}{1.3}$\times 10^{14}$ M$_{\odot}$. This is on par with the $M_{200}$ masses seen in other $z\sim 1.8$ clusters confirmed via grism spectroscopy, e.g. \cite{Stanford2012} reports a mass of  $M_{200} = 5.3\times 10^{14}$ M$_{\odot}$ for a cluster at $z=1.75$ and \cite{Newman2014} finds $M_{200} \simeq (2-3)\times 10^{14}$ M$_{\odot}$ for another $z=1.80$ cluster. 

\subsection{Additional Structure \label{sec:structure}}

\begin{figure}
\centering
\includegraphics[width=\linewidth]{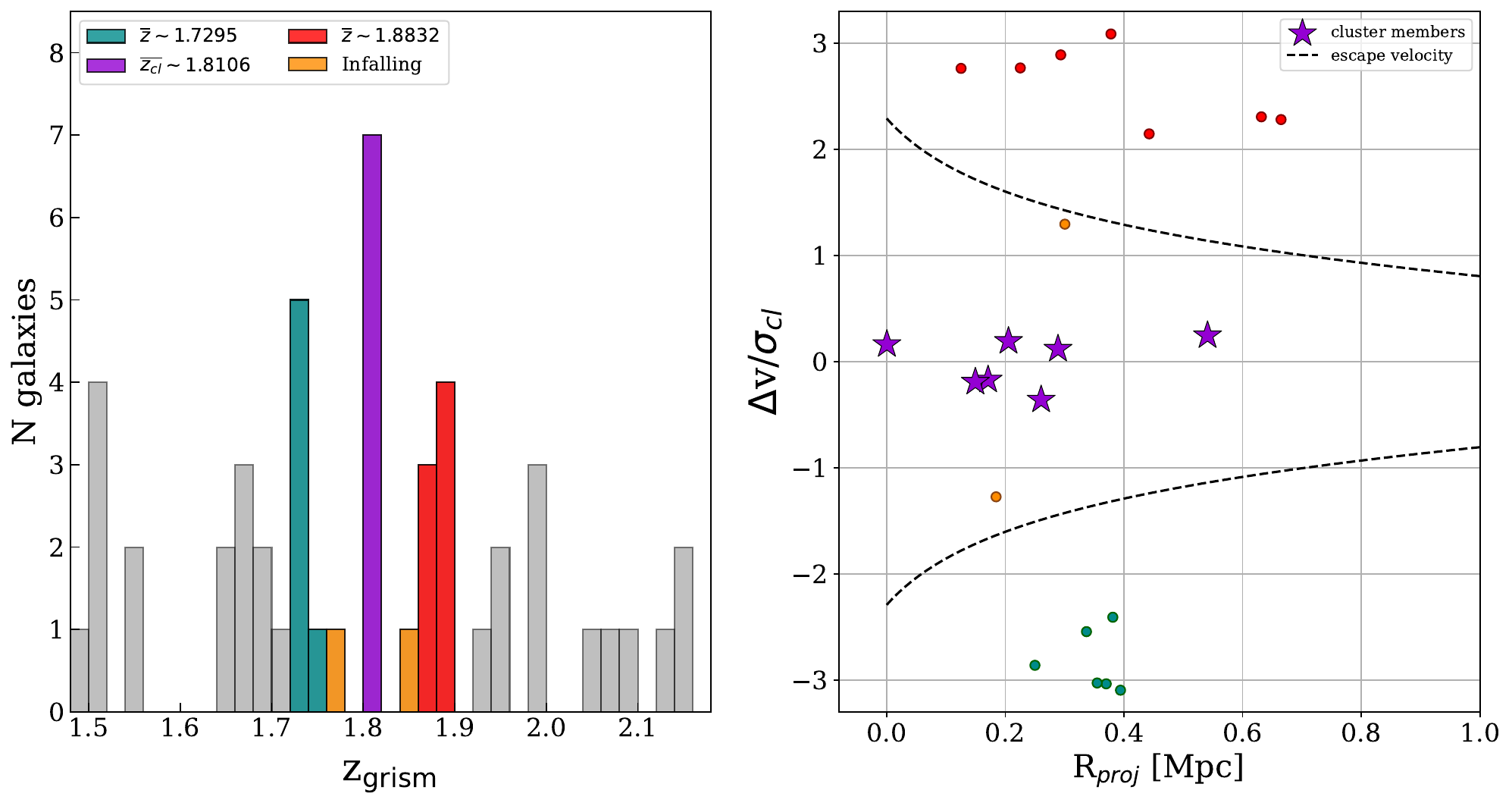}
\caption{\emph{Left:} Redshift histogram of HST sources with bin widths corresponding to 0.02 in redshift, similar to the bottom panel of Fig.\ \ref{fig:zhist} but restricted to the range $1.48 < z_{grism} < 2.18$. \emph{Right:} Normalized projected phase-space diagram for the sources between $1.72 < z_{grism} < 1.89$. Spectroscopically-confirmed members of \targ\ ($z\sim 1.81$) are indicated in purple. The foreground association at $z\sim1.73$ is indicated in teal while the background association at $z\sim 1.88$ is highlighted in red. The two outskirt galaxies that lie between these structures and the \targ, and which appear to be gravitationally bound to the cluster, are highlighted in orange. The dashed black line corresponds to the cluster escape velocity curves (see \S \ref{sec:structure}).} 
\label{fig:phase}
\end{figure}

To identify potential galaxies or substructure which may be infalling into the cluster, we expand our search to focus on the peaks near $z\sim 1.73$ and $z\sim 1.88$ (see left panel of Fig.\ \ref{fig:phase}) and repeat the above analysis. While not associated with the \targ\ structure, we also look at the 11 sources near $z\sim 0.75$ since it is a strong peak in the bottom panel of Fig.\ \ref{fig:zhist} and could be a foreground cluster. We search for sources within $\pm 2,000$ km/s of these redshifts, determine the mean redshift and velocity dispersion for each structure, and then perform the 3$\sigma_{\parallel}$ method described above. Sources that passed the 3$\sigma_{\parallel}$ cleaning were then studied in projected phase-space. 

Following the methods of \cite{Jaffe2015}, we create a phase-space diagram (right panel of Fig.\ \ref{fig:phase}) to visualize where these potential foreground and background structures are with respect to the spectroscopically identified members of \targ. To determine the projected radii, we assume the location of the quasar source represents the cluster center. The peculiar line of sight velocity of each source is calculated as $\Delta v_i = c(z_i - z_{cl})/(1+z_{cl})$ and is normalized by the velocity dispersion of the cluster.

The escape velocity (dashed lines in Fig.\ \ref{fig:phase}), projected to the line of sight, is calculated following the equations in \cite{Jaffe2015}, which assumes a \citet[NFW]{Navarro1996} dark matter halo profile and a fixed concentration parameter $c=6$. The escape velocity is calculated using the virial radius and mass of the main cluster (see \S\ref{sec:clustermass} above). The escape velocity curves define the boundary of any potentially virialized structure. All seven spectroscopically identified members remain within the escape velocity curves, as expected. However, there are two sources (highlighted in orange in Fig.\ \ref{fig:phase}) that fall just within the escape velocity bounds suggesting they could be infalling galaxies gravitationally bound to the \targ\ structure. These sources are at redshifts of $z_{grism} = 1.7751 \pm 0.0460$ and $z_{grism} = 1.8467\pm 0.0010$.

The spatial distribution of the \targ\ confirmed members and these additional structures, at $z\sim 0.75, 1.73$ and $1.88$, is discussed further in \S \ref{sec:spatialdist}.

\subsubsection{Foreground and Background Structures at $z\sim 1.73$ and $z\sim 1.88$}

The foreground structure (highlighted teal in Fig.\ \ref{fig:phase}) consists of 6 sources spanning $1.7244 \leq z_{grism} \leq 1.7435$, with a mean redshift of $\langle z \rangle = 1.7318 \pm 0.0073$ and velocity dispersion $\sigma_{\parallel} =$ \fe{879}{511}{184} km/s. With this, we estimate a virial radius of $R_{200} = $ \fe{0.83}{0.48}{0.17} Mpc and corresponding virial mass of $M_{200} = $ \fe{4.5}{7.8}{2.8} $\times 10^{14} \text{ M}_{\odot}$.

The background structure (highlighted red in Fig.\ \ref{fig:phase}) consists of 7 sources spanning $1.8704 \leq z_{grism} \leq 1.8966$, with a mean redshift $\langle z \rangle = 1.8832 \pm 0.0092$ and velocity dispersion $\sigma_{\parallel} =$ \fe{1038}{514}{205} km/s. After performing the 3$\sigma_{\parallel}$ cleaning, no sources were rejected in either structure. With this, we estimate $R_{200} = $ \fe{0.92}{0.45}{0.18} Mpc and $M_{200} = $ \fe{6.9}{10.2}{4.1} $\times 10^{14} \text{ M}_{\odot}$.

Based on the locations of these structures in the projected phase-space diagram (Fig.\ \ref{fig:phase}) they are not currently infalling into \targ. However, they could be connected to \targ\ on larger scales. We discuss this more in \S\ \ref{sec:spatialdist}.

\subsubsection{Structure at $z\sim 0.75$}

As mentioned in \S \ref{sec:results}, there appears to be a possible foreground structure near $z\sim0.7$. We find that all 11 sources in the peak near $z\sim 0.75$ (see bottom panel of Fig.\ \ref{fig:zhist}) are within $\pm 2000$ km/s of an assumed median redshift of 0.75. With these 11 sources, the structure has a mean redshift of $\langle z \rangle = 0.7502$, with a deviation of 0.006 and a velocity dispersion of $\sigma_{\parallel} = 1140^{+378}_{-189}$ km/s. Additionally, all 11 sources satisfy the 3$\sigma_{\parallel}$ cut. We also estimate the virial radius and mass using the above equations for this lower-z structure and find $R_{200} = $ \fe{1.86}{0.62}{0.31} Mpc and $M_{200} = 1.7^{+1.7}_{-0.8} \times 10^{15}$ M$_{\odot}$. However, if the structure is still in the process of formation, the assumption of virial equilibrium may not hold. In this unrelaxed state, short-term dynamical instabilities could inflate the observed velocity dispersion, leading to a mass estimate that is higher than the system's true mass.

\subsection{Properties of the Candidate High-Redshift Cluster \targ \label{sec:properties}}

\subsubsection{Photometric Properties}

In the optical bands, five candidate members are observed in both the $r-$ and $i-$imaging (Table \ref{tab: candmembers}). However, for four of these members in the $i-$band and two of these members in the $r$-band, their magnitudes are above the limits defined in \S \ref{sec:addphot} and thus we do not include these points in the SED fitting of these sources. The LDT r- and i-band observations have limiting AB magnitudes of 25 and 24, respectively. Thus, we include these as upper limits on the optical magnitudes in Table \ref{tab: candmembers}. For the quasar source (ID 310) the large offset between the observed optical photometry and the best-fit model seen in Fig.\ \ref{fig:candfits}, is due to the excess optical emission of the quasar source, as the FSPS templates used in the \griz\  fits only model assumed stellar contributions to the SED.

Of our seven candidate members, two (IDs 310 \& 470) are detected in both the 3.6 \micron\ and 4.5 \micron\ Spitzer images. However, source 470's magnitudes (see Table \ref{tab: candmembers}) are fainter than the $m_{3.6},m_{4.5}\leq 21.4$ limit we define in \S \ref{sec:addphot} and we therefore do not include these points in the SED fitting of this source. 

\begin{figure}
\begin{center}$
\begin{array}{c}
\includegraphics[width = 0.45 \textwidth]{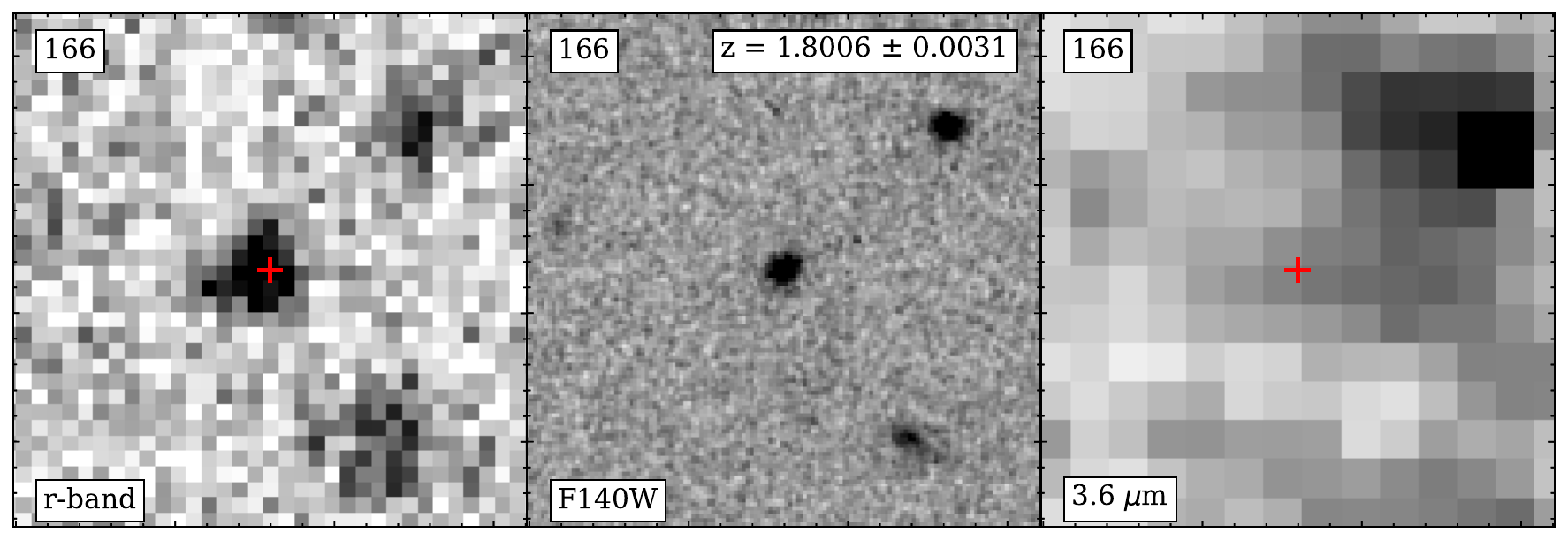} \\
\includegraphics[width = 0.45 \textwidth]{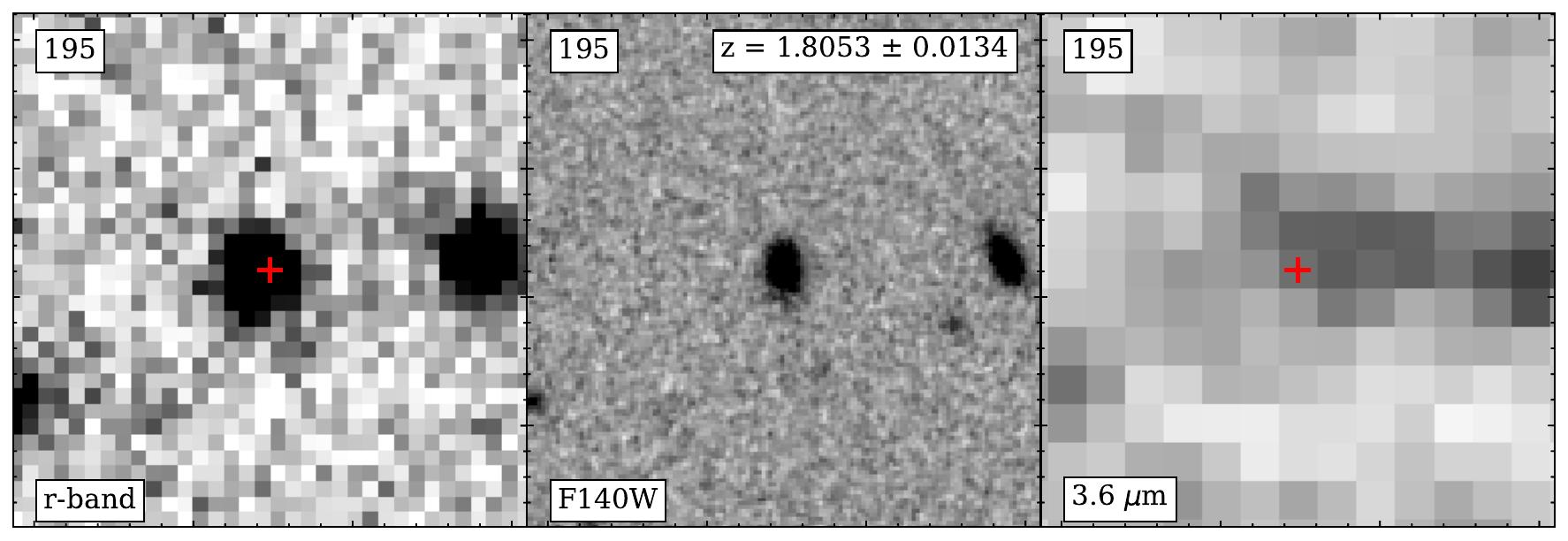} \\
\includegraphics[width = 0.45 \textwidth]{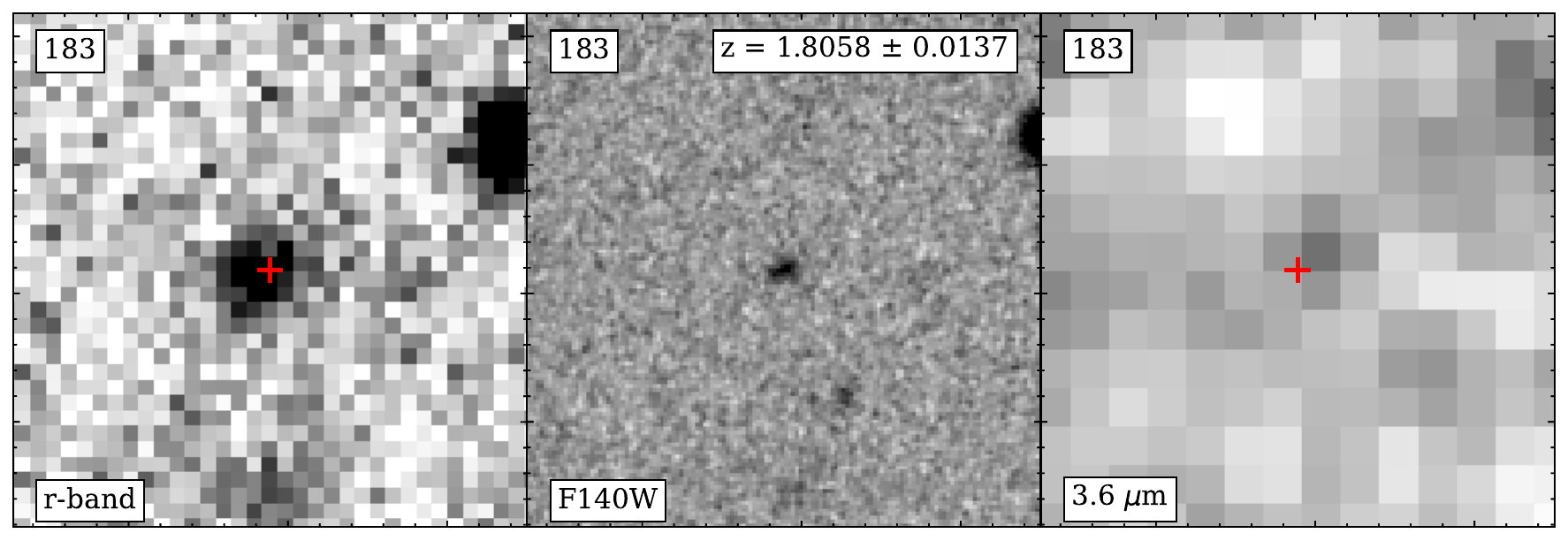} \\
\includegraphics[width = 0.45 \textwidth]{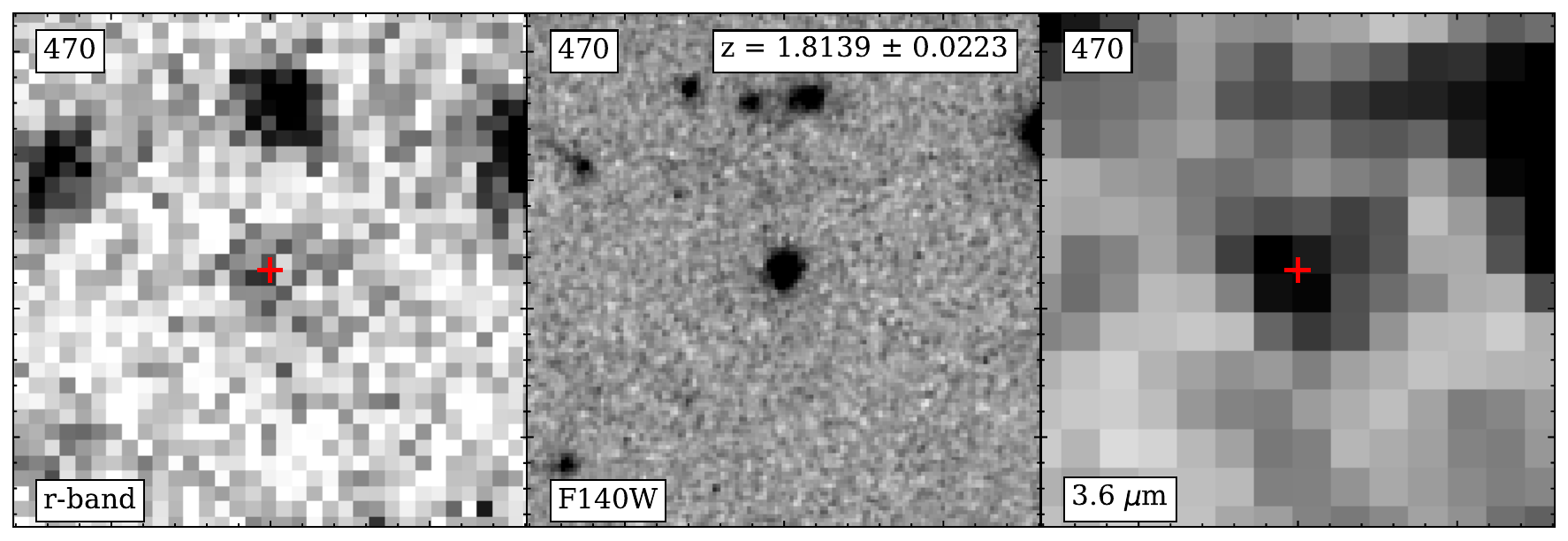} \\
\includegraphics[width = 0.45 \textwidth]{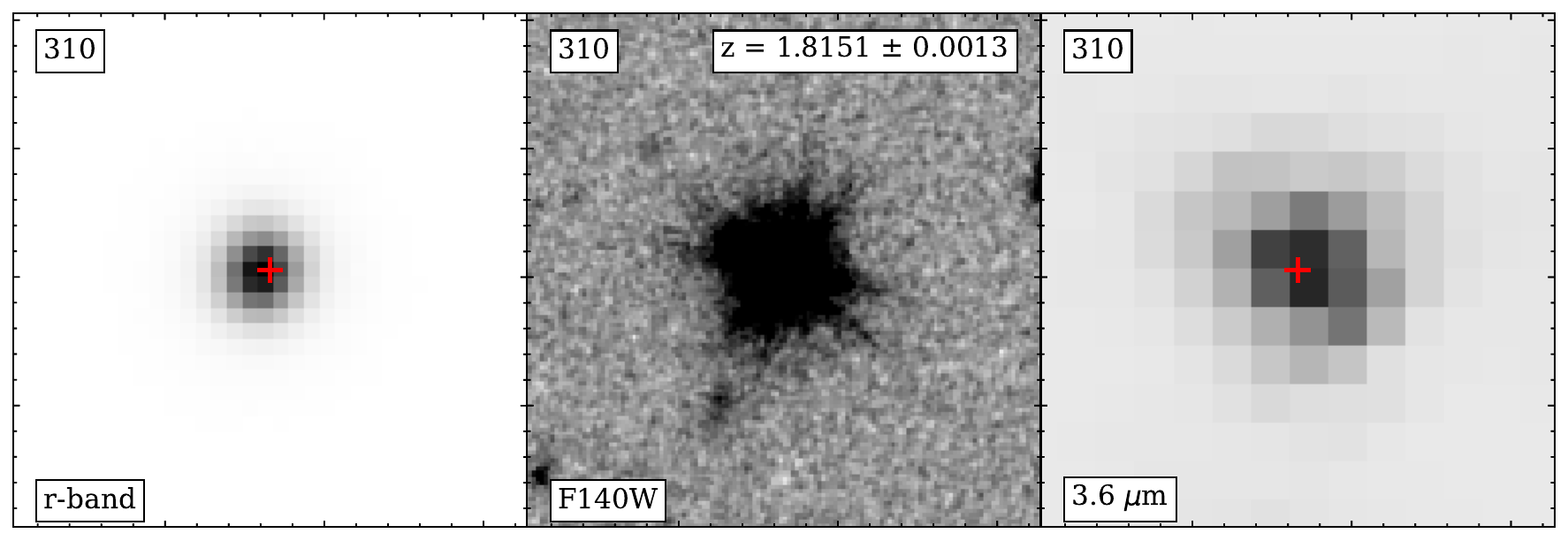} \\
\includegraphics[width = 0.45 \textwidth]{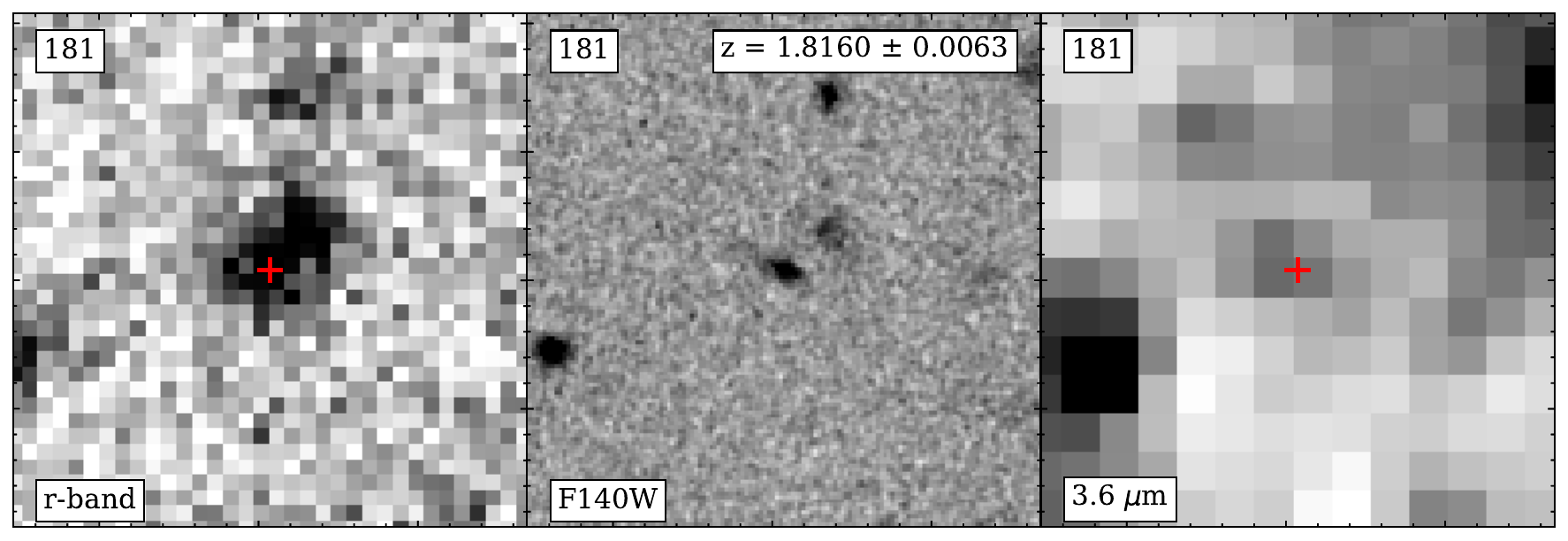} \\ 
\includegraphics[width = 0.45 \textwidth]{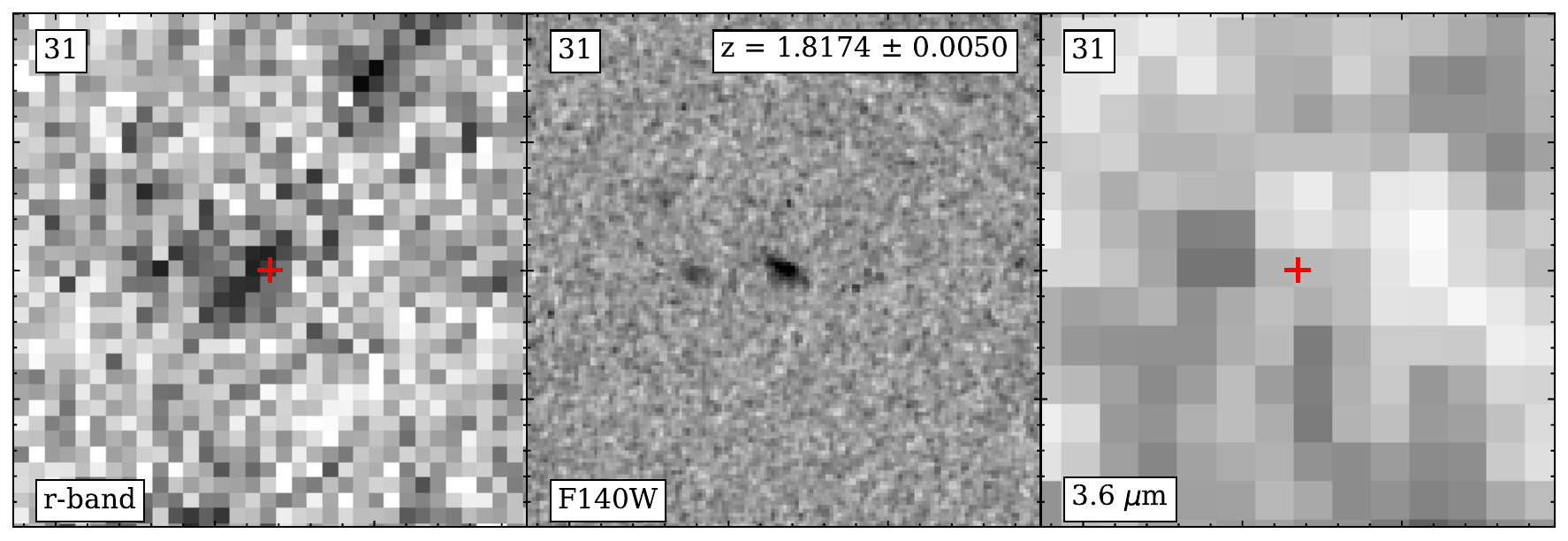} \\ 
\end{array}$
\caption{Thumbnail images (8\arcsec$\times$8\arcsec) of the spectroscopically identified members of the candidate cluster \targ\ from LDT r-band (left; \citetalias{GM2019}), HST F140W (center), and Spitzer 3.6 \micron\ (right; \citetalias{PM2017}). The red cross overlaid on the LDT and Spitzer thumbnails corresponds to the centroid of the source in the HST image. \label{fig:thumbs}}
\end{center}
\end{figure}

Figure \ref{fig:thumbs} shows 8\arcsec$\times$8\arcsec\ thumbnails for each of these seven sources as observed in the optical (LDT r-band; left panel), near-infrared (HST F140W; middle panel), and mid-infrared (Spitzer 3.6 \micron; right panel). 

\subsubsection{Star-Formation Rates}

\griz\ can be used to compute stellar population synthesis (SPS) parameters (such as $L_{\nu}$, $M_{*}$, and star formation rate (SFR)) from the fitted FSPS templates by normalizing the fit coefficients (which dictate how much each template contributes to the best-fit model) to the V-band luminosity, and scaling the template-derived SFRs using the galaxy's observed luminosity and redshift. Table \ref{tab:fitinfo} lists the SFR and stellar mass as determined by the \griz\ best-fit templates. Stellar masses and SFRs derived from fits to FSPS templates assume a \cite{Chabrier2003} initial mass function (IMF). Excluding the quasar source, because the SFR value is likely overestimated due to contamination from AGN emission, the \griz\ estimated SFRs for candidate \targ\ members range between 3-9 M$_{\odot}$/yr with a cluster mean of about 5 $\pm$ 9 M$_{\odot}$/yr for the main seven members. However, since in the case of our seven candidate members, these SFRs are estimated from either the G141 grism data alone or with just 1-2 additional photometric data points, they are not well constrained and additional observations over a wider range of wavelengths are needed for more robust modeling of the SEDs using the limited suite of FSPS templates. While the \griz\ derived values are reported here for completeness, we perform an additional calculation of the SFRs by utilizing the fact that all candidate members have spectra exhibiting [OIII]$\lambda5007$ emission lines, indicating active or on-going star formation should be occurring, and therefore should exhibit SFRs $>6$ M$_{\odot}$/yr \citep{Daddi2007,Rodighiero2011}. To do this, we use the [OIII] line flux measured using \griz\ during the second pass of grism fitting where individual emission lines were fit (see \S\ref{sec:grizfit}). Following methods similar to \cite{Noirot2016} and \cite{Noirot2018}, we use the \cite{Kennicutt1983} relation $SFR = L(H\alpha)/(1.12\times 10^{41}\ \text{ergs/s})$, which assumes a \cite{Salpeter1955} IMF. We assume a line ratio of [OIII]/H$_{\alpha}\sim 1$, add a 0.35 dex scatter to the errors on the [OIII] line fluxes to account for the observed scatter in [OIII]/H$_{\alpha}$, and implement a rough correction for dust attenuation\footnote{The [OIII] flux already has a correction for foreground extinction from \griz\ using the \cite{Fitzpatrick1999} law however an additional correction for the galaxy's internal dust attenuation must applied.}$^,$\footnote{assuming a constant $A_V = 1$ mag in the V-band, typical for star-forming galaxies \citep{Kewley2004}, and use the \cite{Calzetti2000} attenuation curve with $R_V = 4.05$.}. 

The dust corrected SFR$_{\text{[OIII]}}$ estimates for the candidate cluster members are listed in Table \ref{tab:fitinfo}. The \griz\ derived SFRs can be converted from \cite{Chabrier2003} IMF to the \cite{Salpeter1955} IMF by multiplying by a constant factor of 1.7 \citep{Genzel2010,Madau2014,Speagle2014,Figueira2022}. However, for ease of comparison with other studies, we stick to the SFRs as calculated by the \cite{Kennicutt1983} relation, which assumes a \cite{Salpeter1955} IMF. Excluding the quasar source as the SFR is likely to be overestimated due to contamination from AGN narrow-line emission regions, the [OIII] SFRs of our candidate members range from 11-43 M$_{\odot}$/yr, with the cluster mean of $\sim$ 25 $\pm$ 7 M$_{\odot}$/yr for the main seven members. This is consistent with what has been previously seen in clusters at similar redshifts. \cite{Noirot2016} and \cite{Noirot2018} find a mean SFR of $\sim 40$ M$_{\odot}$/yr, determined from [OIII] line fluxes, for star-forming members of nine $z\sim 1.9$ CARLA structures. For XMM-LSS J02182-05102, a cluster at $z=1.6$, \cite{Tran2015} determine the SFR of cluster members based on H$\alpha$ line fluxes and show an average rate of 20-22 M$_{\odot}$/yr (converted to a \cite{Salpeter1955} IMF), but they note that the average SFR per H$\alpha$-detected galaxy in the cluster core is about half that measured at larger radii ($> 1$ Mpc). However, other clusters at similar redshifts either show an extreme central starburst --- e.g. SpARCS1049$+$56 $>800$ M$_{\odot}$/yr at $z=1.7$ \citep{Webb2015} --- or are largely quiescent, as in JKCS041 at $z=1.8$ \citep{Newman2014} which, in contrast to \targ, has only a handful of spectroscopically confirmed SF members.

Based on the stellar mass-SFR relation of \cite{Daddi2007}, the SFRs for four of the seven candidate members place them in the star-forming main sequence at $1.5 \leq z \leq 2.5$, denoting them as actively star-forming galaxies.  

\begin{figure}
 % \figurenum{8}
    \centering
\plotone{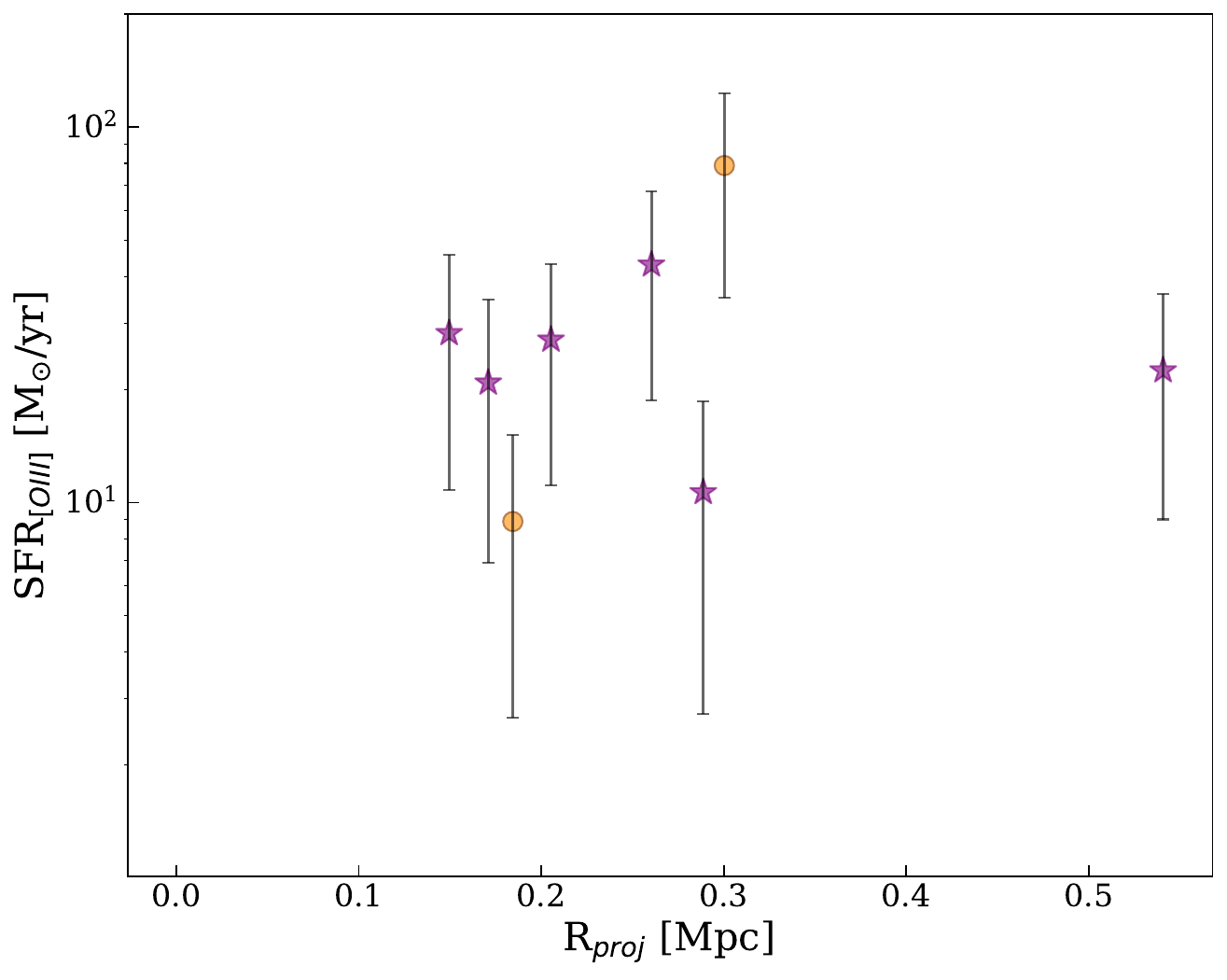}
    \caption{Star formation rate (SFR) versus projected cluster-centric distance for sources identified as spectroscopic members of \targ\ (purple stars) and two potential infalling members (orange circles). The SFRs are determined from [OIII] emission line flux and assume a \cite{Salpeter1955} IMF. The cluster center is defined as the location of the quasar source, which is not shown here as its SFR is likely overestimated due to AGN contributions.}
    \label{fig:sfr_vs_rproj}
\end{figure}

To search for possible signs of feedback or quenching, we plot the SFR as a function of projected distance from the cluster center in Fig.\ \ref{fig:sfr_vs_rproj}. The quasar source is not shown as the SFR is likely overestimated due to AGN contributions. We do not observe a significant correlation between SFR and distance from our assumed cluster center.

\subsubsection{Spatial Distribution of Member Sources \label{sec:spatialdist}}

\begin{figure*}
 \epsscale{0.7}
    \centering
\plotone{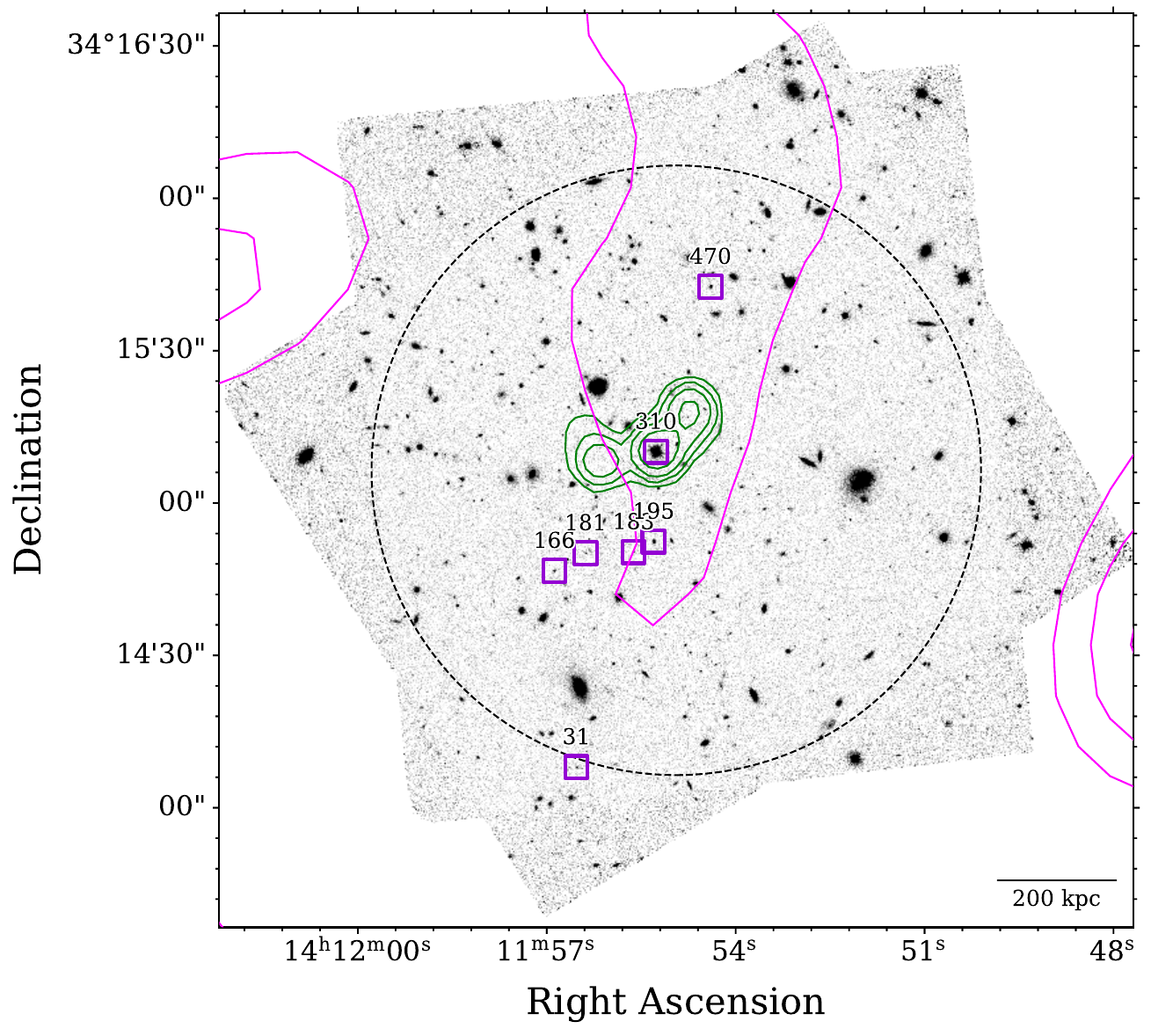}
    \caption{Spatial distribution of candidate cluster members (purple squares) in \targ. Source IDs correspond to those listed in Tables \ref{tab: candmembers} \& \ref{tab:fitinfo}. Overlaid are the VLA FIRST radio contours (green) of the bent, double-lobed radio source, which initially flagged \targ\ as a high-z cluster candidate. Centered on the bent radio AGN is a 1\arcmin\ dashed, black circle. Magenta contours show the projected density of red (high-z candidate) galaxies as measured by \citetalias{GM2019}. The scale bar in the lower right corner is defined at the redshift of the candidate cluster ($z=1.8$).}
    \label{fig:cands_spatial}
\end{figure*}

Figure\ \ref{fig:cands_spatial} shows the spatial distribution of our seven spectroscopically identified members in the HST imaging. Overlaid are contours of the projected density of red galaxies as measured by \citetalias{GM2019} (magenta) and the 20 cm radio emission (green) from the bent radio AGN. The spatial distribution of our seven candidate cluster members follows roughly the distribution of red members identified in \citetalias{GM2019} (see magenta contours of Fig. \ref{fig:cands_spatial}), forming an elongated structure along the N-S axis. 

\begin{figure}
    \centering
\plotone{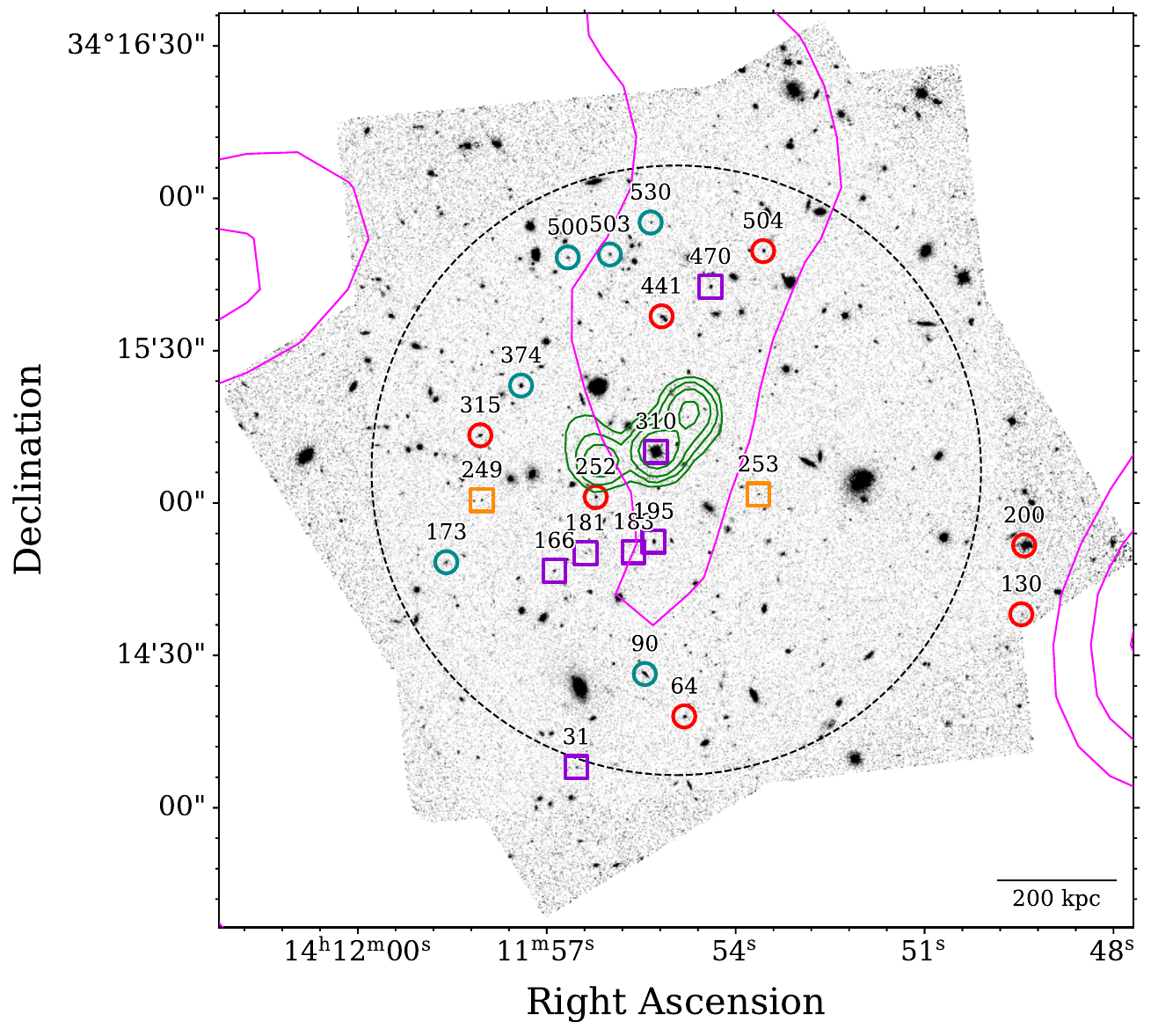}
    \caption{Spatial distribution of sources in the foreground structure at $\langle z \rangle = 1.7318$ (teal circles) and in the background structure at $\langle z \rangle = 1.8832$ (red circles) with respect to the spectroscopically identified cluster at $\langle z \rangle = 1.8106$, \targ\ (purple squares) and the two potentially infalling members (orange squares). The contours and scale bar are as defined in Fig.\ \ref{fig:cands_spatial}.}
    \label{fig:forespatial1}
\end{figure}

Nearly all of the spectroscopically identified \targ\ members lie within $\sim$1\arcmin\ ($\sim0.5$ Mpc at $z=1.8151$) of the quasar source, denoted by the large dashed circle in Figs.\ \ref{fig:rgb} \& \ref{fig:cands_spatial}, satisfying the cluster membership criteria set forth in \cite{Noirot2018}. There is one source (ID 31) that lies just outside this radius, sitting at 1.07\arcmin\ ($\approx 0.54$ Mpc) from the quasar source. However, the quasar may not represent the true cluster center and we therefore choose to include this source in the cluster membership. The two additional, possibly infalling sources, identified in the phase-space analysis (see \S\ref{sec:structure}), are highlighted by orange squares in Fig.\  \ref{fig:forespatial1}.

Notably, the bent, double-lobed radio source, which exhibits a wide-angle tail (WAT) morphology, also follows the N-S trend as the radio lobes are bent towards the north. The tails of the WAT radio sources are bent via ram-pressure arising from relative motions between the host galaxy and an external medium --- in this case the potential ICM of \targ. Previous studies have shown that these bent-lobed radio morphologies can be the result of cluster merger activity \citep{Roettiger1993, Bliton1998, Douglass2011, Mendygral2012, PM2013, Golden-Marx2021, Watson2023}. Thus, the cluster could be in an unrelaxed state, which would increase the measured velocity dispersion. The mass we calculate assuming virialization should then be considered an upper limit to the true cluster mass. 

The spatial distribution of sources in the additional fore- and background structures identified at $z\sim 1.73$ and $z\sim 1.88$ (see \S \ref{sec:structure}) are shown in Fig.\ \ref{fig:forespatial1}, in teal and red circles, respectively. The similar spatial distributions and small separations --- both in projected space (within $\sim$1\arcmin) and redshift ($\Delta z \sim 0.08$) --- relative to \targ, suggest they could be a part of the same large-scale filament. Together, the three structures may be a protocluster system that would eventually merge and form a single, massive cluster.

\begin{figure}
    \centering
\plotone{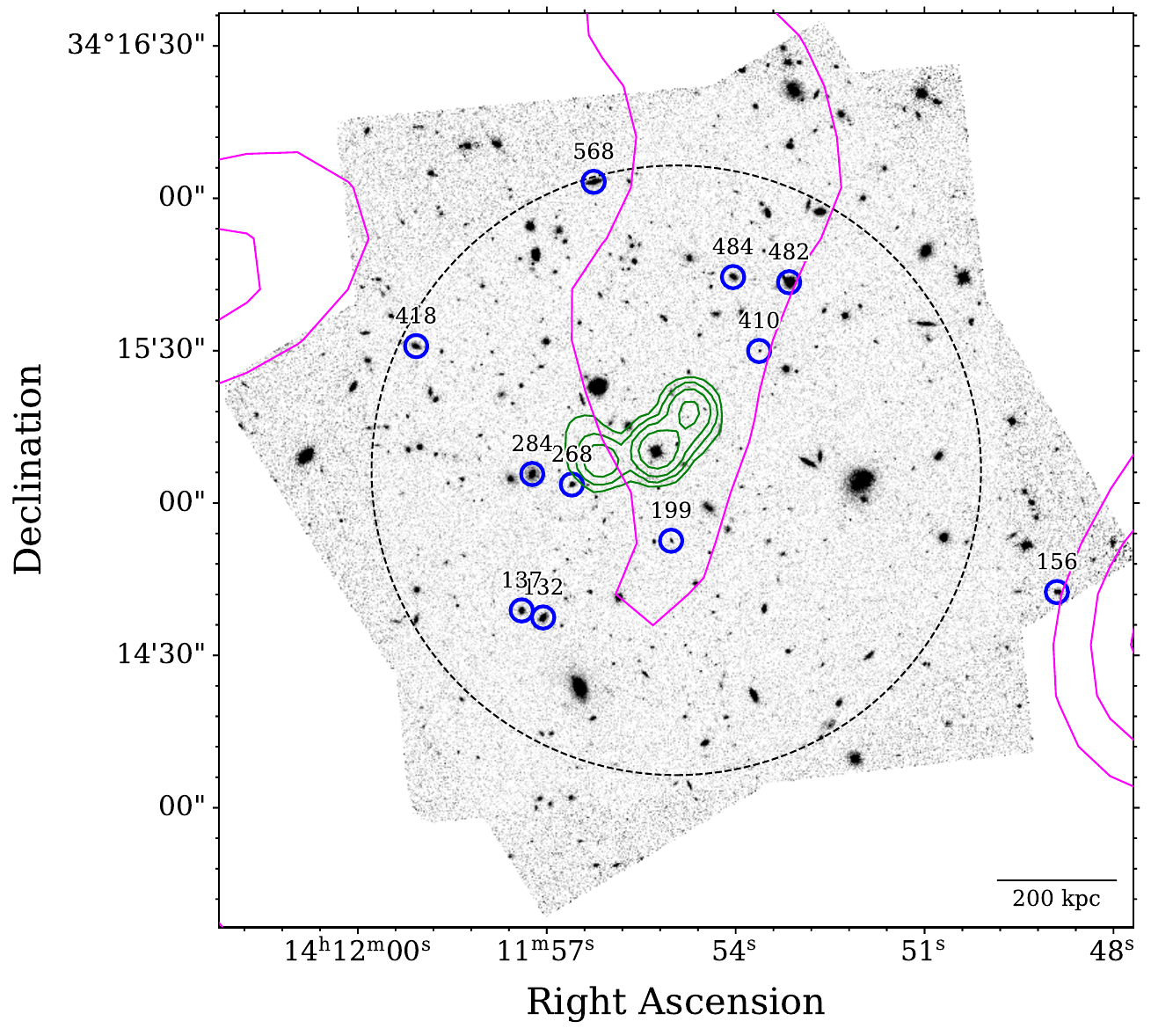}
    \caption{Spatial distribution of members (blue circles) in the foreground structure at $\langle z \rangle = 0.7502$. The contours and scale bar are as defined in Fig.\ \ref{fig:cands_spatial}.}
    \label{fig:forespatial2}
\end{figure}

We also show the spatial distribution of sources in the foreground structure at $\langle z \rangle = 0.7502$ in Fig.\ \ref{fig:forespatial2}, however, the sources show no notable distribution. 

\section{Summary \& Conclusions \label{sec:concl}}

\targ\ was initially identified as a high-z cluster candidate in the COBRA survey using a bent, double-lobed radio source as a signpost. Using Spitzer/IRAC observations, \citetalias{PM2017} measure a 3$\sigma$ significant overdensity of red sources within 2\arcmin\ of the bent radio AGN. Building on that work, further analysis of the Spitzer data by \citetalias{GM2019} showed a 2.2$\sigma$ significant overdensity of red-sequence members. As part of a joint Chandra-Hubble proposal (PI E. Blanton), additional follow-up X-ray and NIR observations were taken in order to confirm \targ\ as a high-z structure. Here we presented the results of the HST/WFC3 grism observations. 

We identified seven candidate members within a $\sim$0.5 Mpc radius whose spectroscopic redshifts lie within $\pm 2,000$ km/s, corresponding to a range of $1.7963 \leq z_{grism} \leq 1.8339$, thus spectroscopically confirming \targ\ as a bona-fide high-z structure with a mean redshift of $\langle z_{grism}\rangle = 1.8106 \pm 0.0006$ and an actual z range of $z_{grim} = 1.8006-1.8175$ (or $\pm$ 1804 km/s). The line-of-sight velocity dispersion of \targ\ is found to be $\sigma_{\parallel} = 701^{+347}_{-138}$ km/s. From this, we calculate the virial radius of the cluster as $R_{200} = $\fe{0.64}{0.32}{0.13} Mpc corresponding to a virial mass, $M_{200} \approx 2.2^{+3.3}_{-1.3}\times 10^{14}$ M$_{\odot}$.

We estimate dust-corrected SFRs for individual members using the [OIII] line fluxes. We find that \targ\ has a mean SFR of $\sim 25 \pm 7$ M$_{\odot}$/yr, with over half of the spectroscopically identified cluster members having SFRs consistent with that of the star-forming main sequence at redshifts $1.5\leq z\leq 2.5$. 

Spatially, the spectroscopic members of \targ\ form an elongated N-S structure that appears aligned with the distribution of red-sequence members seen in \citetalias{GM2019}. Interestingly, the bent, double-lobed radio source, which exhibits a wide-angle tail (WAT) morphology, also follows this N-S trend, as the radio lobes are bent towards the north. This could suggest previous or on-going merger activity with another galaxy cluster or group infalling along that axis. Thus, the cluster may not be virialized and may still be in formation.

Looking in the projected phase-space, we identified two additional sources which could be infalling galaxies in the cluster outskirts. We also identify structures, at $z\sim 1.73$ and $z\sim1.88$, that could be connected to \targ\ via a large-scale filament. These two additional structures, along with \targ, could represent components of a larger-scale protocluster system.

With our relatively shallow HST grism observations, we are biased towards star-forming galaxies, with their easily identified emission lines, and are likely missing out on a population of passive, early-type galaxies. Additional, deeper grism observations would help in identifying these sources, thereby increasing the identified cluster population and providing a more accurate estimate of the cluster's velocity dispersion and mass. Additional photometric observations would also allow for a more robust analysis of galaxy SEDs, leading to a more accurate estimate of galaxy star-formation rates and masses.

At $\langle z_{grism}\rangle = 1.8106 \pm 0.0006$, \targ\ is the highest redshift cluster to be spectroscopically confirmed using a bent, double-lobed radio source as a cluster signpost. It underscores the utility of using such radio sources to discover clusters at a wide range of redshifts.

\begin{deluxetable*}{ccccc}
\tablecaption{Coordinates, HST photometry, and grism redshifts for all sources with successful fits in the HST observations.\label{tab:allfits}}
\tablehead{\colhead{ID}  & \colhead{RA}  & \colhead{DEC}  & \colhead{$m_{F140W}$}  & \colhead{$z_{grism}$} }
\startdata
15 & 14:11:55.72 & 34:13:57.76 & $24.17 \pm 0.07$ & $0.9275 \pm 0.0059$\\
19 & 14:11:57.17 & 34:14:00.05 & $22.84 \pm 0.03$ & $0.3842 \pm 0.0124$\\
20 & 14:11:56.95 & 34:14:00.76 & $22.73 \pm 0.03$ & $1.1882 \pm 0.0048$\\
22 & 14:11:57.11 & 34:14:01.83 & $22.26 \pm 0.02$ & $1.2808 \pm 0.0055$\\
23 & 14:11:56.61 & 34:14:02.03 & $20.60 \pm 0.00$ & $0.1109 \pm 0.0173$\\
26 & 14:11:54.96 & 34:14:04.77 & $23.67 \pm 0.06$ & $2.2464 \pm 0.0527$\\
\multicolumn{5}{c}{$\dots$} \\
\enddata
\tablecomments{Table \ref{tab:allfits} is published in its entirety in the machine-readable format.
      A portion is shown here for guidance regarding its form and content.}
\end{deluxetable*}

\begin{deluxetable*}{cccccccc}
\tablecaption{Coordinate and photometric information for spectroscopically identified members of \targ\ and the two possible infalling members. All magnitudes are reported in the AB system.  
\label{tab: candmembers}}
\tablehead{
    \colhead{ID}& 
    \colhead{RA}& 
    \colhead{DEC}& 
    \colhead{$m_{\text{F140W}}$}& 
    \colhead{$m_{3.6}$\tablenotemark{a}}&
    \colhead{$m_{4.5}$\tablenotemark{a}}& 
    \colhead{$m_r$\tablenotemark{b}}& 
    \colhead{$m_i$\tablenotemark{b}}
    }
\startdata
31 & 14:11:56.53 & 34:14:08.01 & $24.49 \pm 0.08$ & $> 22.15$\tablenotemark{c} & $> 22.5$\tablenotemark{c} & $25.90 \pm 0.12$\tablenotemark{d} & $26.82 \pm 0.23$\tablenotemark{d} \\
166 & 14:11:56.88 & 34:14:46.67 & $23.57 \pm 0.04$ & $> 22.15$\tablenotemark{c} & $> 22.5$\tablenotemark{c} & $24.94 \pm 0.07$ & $24.90 \pm 0.11$\tablenotemark{d} \\
181 & 14:11:56.39 & 34:14:50.16 & $24.35 \pm 0.08$ & $> 22.15$\tablenotemark{c} & $> 22.5$\tablenotemark{c} & $>25$\tablenotemark{c} & $>24$\tablenotemark{c}\\
183 & 14:11:55.62 & 34:14:50.36 & $24.80 \pm 0.10$ & $> 22.15$\tablenotemark{c} & $> 22.5$\tablenotemark{c} & $>25$\tablenotemark{c} & $>24$\tablenotemark{c}\\
195 & 14:11:55.30 & 34:14:52.42 & $22.98 \pm 0.03$ & $> 22.15$\tablenotemark{c} & $>22.5$\tablenotemark{c}  & $24.25 \pm 0.04$ & $24.52 \pm 0.08$\tablenotemark{d}\\
310\tablenotemark{e} & 14:11:55.27 & 34:15:10.11 & $18.46 \pm 0.00$ & $18.12 \pm 0.01$ & $17.79 \pm 0.01$  & $18.33 \pm 0.01$ & $18.21 \pm 0.01$\\
470 & 14:11:54.40 & 34:15:42.60 & $23.02 \pm 0.03$ & $22.38 \pm 0.17$\tablenotemark{d} & $23.39 \pm 0.30$ \tablenotemark{d}  & $27.47 \pm 0.22$\tablenotemark{d} & $25.31 \pm 0.13$\tablenotemark{d}\\ \hline
249 & 14:11:58.03 & 34:15:00.58 & $24.56 \pm 0.08$ & $> 22.15$\tablenotemark{c} & $> 22.5$\tablenotemark{c}  & $25.65 \pm 0.10$\tablenotemark{d} & $> 24$\tablenotemark{c}\\
253 & 14:11:53.64 & 34:15:01.69 & $24.36 \pm 0.07$ & $> 22.15$\tablenotemark{c} & $> 22.5$\tablenotemark{c}  & $> 25$\tablenotemark{c} & $25.37 \pm 0.14$\tablenotemark{d}
\enddata
\tablenotetext{a}{Spitzer 3.6 and 4.5 \micron\ photometry from \cite{PM2017}}
\tablenotetext{b}{LDT r- and i-band photometry from \cite{GM2019}}
\tablenotetext{c}{Was not detected in the image and therefore represents an upper limit on the magnitude}
\tablenotetext{d}{Does not satisfy our defined magnitude criteria ($m_{3.6},m_{4.5}\leq 21.4$, $m_i \leq 24$, and $m_r \leq 25$;  see \S \ref{sec:addphot} \& \S\ref{sec:grizfit}) and while these measurements are not used in the redshift fitting for these sources, we include them here for completeness }
\tablenotetext{e}{The quasar hosting the bent, double-lobed radio source which identified \targ\ as a high-z cluster candidate }
\end{deluxetable*}

\begin{deluxetable*}{cccccc}
\tablecaption{Spectroscopically identified members of \targ\ and the two possible infalling members. \label{tab:fitinfo} }
\tablehead{
    \colhead{ID}& 
    \colhead{Redshift\tablenotemark{a}}&
    \colhead{$\Delta v$}&
    \colhead{Mass\tablenotemark{b}}& 
    \colhead{SFR\tablenotemark{b}}&
    \colhead{SFR$_{\text{[OIII]}}$\tablenotemark{c}}\\
    \colhead{}& 
    \colhead{}&
    \colhead{[km\ s$^{-1}$]}&
    \colhead{[M$_{\odot}$]}&
    \colhead{[M$_{\odot}$\ yr$^{-1}$]} &
    \colhead{[M$_{\odot}$\ yr$^{-1}$]} 
    }
\startdata
31 & $1.8174 \pm 0.0050$ & 737 & $2.0\times 10^9$ & $4 \pm 16$ & $22 \pm 13$\\
166 & $1.8006 \pm 0.0031$ & -1067 & $9.4\times 10^9$ & $5 \pm 3$ & $43 \pm 24$\\
181 & $1.8160 \pm 0.0063$ & 576 & $2.5\times 10^{10}$ & $5 \pm 2$ & $27 \pm 16$\\
183 & $1.8058 \pm 0.0137$ & -512 & $6.3\times 10^8$ & $3 \pm 1$ & $21 \pm 14$\\
195 & $1.8053 \pm 0.0134$ & -566 & $7.6\times 10^9$ & $9 \pm 57$ & $28 \pm 17$\\
310\tablenotemark{d} & $1.8151 \pm 0.0013$ & 480 & $5.8\times 10^{11}$ & \dots & \dots\\
470 & $1.8139 \pm 0.0223$ & 352 & $1.5\times 10^{10}$ & $6 \pm 4$ & $11 \pm 8$\\ \hline
249 & $1.8467 \pm 0.0010$ & 3853 & $2.0\times 10^{10}$ & $26 \pm 34$ & $79 \pm 44$\\
253 & $1.7751 \pm 0.0460$ & -3789 & $1.2\times 10^{10}$ & $4 \pm 86$ & $9 \pm 6$\\
\enddata
\tablenotetext{a}{\griz\ best-fit redshift. Reported errors are $1\sigma$.}
\tablenotetext{b}{From \griz\ best-fit FSPS templates, assumes \cite{Chabrier2003} initial mass function}
\tablenotetext{c}{Calculated from the dust-corrected [OIII] line flux assuming an H$\alpha$/[OIII] line ratio of unity (see \S \ref{sec:properties}); From a \cite{Salpeter1955} IMF which can be divided by a factor of 1.7 to convert to a \cite{Chabrier2003} IMF}
\tablenotetext{d}{The quasar hosting the bent, double-lobed radio source which identified \targ\ as a high-z cluster candidate}
\end{deluxetable*}

\section*{Acknowledgments}
    
    We thank the anonymous referee for their useful comments and suggestions that greatly helped improve this manuscript. C.B.W thanks Ga\"el Noirot and Gabe Brammer for providing helpful discussion on the use of \griz\ and its data products.
     
     This research is based on observations made with the NASA/ESA Hubble Space Telescope obtained from the Space Telescope Science Institute, which is operated by the Association of Universities for Research in Astronomy, Inc., under NASA contract NAS 5-26555. These observations are associated with program 15994 and were supported by NASA grant HST-GO-15994.001-A. Additional support for this work was provided by NASA through Chandra Award Number GO0-21123X.

     C.B.W. and E.L.B. received partial support through NASA from Chandra Award Number GO2-23123X. C.B.W. was also supported by Massachusetts Space Grant Consortium Award 80NSSC20M0048 to MIT and awards 580379, 633061, 699842, 742790, 793224, 858617, and 908470, to C.B.W at BU.
     
    These results made use of the Lowell Discovery Telescope at Lowell Observatory. Lowell is a private, nonprofit institution dedicated to astrophysical research and public appreciation of astronomy and operates the LDT in partnership with Boston University, the University of Maryland, the University of Toledo, Northern Arizona University, and Yale University. LMI construction was supported by Grant No. AST-1005313 from the National Science Foundation.
    
    This work is based in part on observations made with the Spitzer Space Telescope, which is operated by the Jet Propulsion Laboratory, California Institute of Technology under a contract with NASA. Support for this work was provided by NASA through an award issued by JPL/Caltech (NASA Award RSA No. 1440385).
    
    Funding for SDSS-III has been provided by the Alfred P. Sloan Foundation, the Participating Institutions, the National Science Foundation, and the U.S. Department of Energy Office of Science. The SDSS-III website is http://www.sdss3.org/.
    
    SDSS-III is managed by the Astrophysical Research Consortium for the Participating Institutions of the SDSS-III Collaboration including the University of Arizona, the Brazilian Participation Group, Brookhaven National Laboratory, Carnegie Mellon University, University of Florida, the French Participation Group, the German Participation Group, Harvard University, the Instituto de Astrofisica de Canarias, the Michigan State/Notre Dame/JINA Participation Group, Johns Hopkins University, Lawrence Berkeley National Laboratory, Max Planck Institute for Astrophysics, Max Planck Institute for Extraterrestrial Physics, New Mexico State University, New York University, Ohio State University, Pennsylvania State University, University of Portsmouth, Princeton University, the Spanish Participation Group, University of Tokyo, University of Utah, Vanderbilt University, University of Virginia, University of Washington, and Yale University

    The HST data presented in this article were obtained from the Mikulski Archive for Space Telescopes (MAST) at the Space Telescope Science Institute. The specific observations analyzed can be accessed via \dataset[doi: 10.17909/vmh1-fb56]{https://doi.org/10.17909/vmh1-fb56}.
% \end{acknowledgments}

\bibliography{main.bib}

\end{document}